\documentclass[aps,prl,twocolumn,amsmath,amssymb,nofootinbib,superscriptaddress,floatfix,reprint,longbibliography]{revtex4-2}

\usepackage{graphicx} % 去掉 dvips 选项
\usepackage{bm}
\usepackage{xcolor} % 替换 color 包
\usepackage{float}
\usepackage{url}
\usepackage{ulem}
\usepackage{xpatch}

\usepackage{hyperref}
\usepackage{nameref}
\hypersetup{colorlinks=true, urlcolor=blue, linkcolor=blue, citecolor=blue}

\begin{document}
	\newcommand{\fig}[2]{\includegraphics[width=#1]{#2}}
	\newcommand{\la}{{\langle}}
	\newcommand{\ra}{{\rangle}}
	\newcommand{\dg}{{\dagger}}
	\newcommand{\upa}{{\uparrow}}
	\newcommand{\dna}{{\downarrow}}
	\newcommand{\ab}{{\alpha\beta}}
	\newcommand{\ias}{{i\alpha\sigma}}
	\newcommand{\ibs}{{i\beta\sigma}}
	\newcommand{\hH}{\hat{H}}
	\newcommand{\hn}{\hat{n}}
	\newcommand{\hc}{{\hat{\chi}}}
	\newcommand{\hU}{{\hat{U}}}
	\newcommand{\hV}{{\hat{V}}}
	\newcommand{\br}{{\bf r}}
	\newcommand{\bk}{{{\bf k}}}
	\newcommand{\bq}{{{\bf q}}}
	\def\gsim{~\rlap{$>$}{\lower 1.0ex\hbox{$\sim$}}}
	\setlength{\unitlength}{1mm}
	\newcommand{{\vhf}}{$\chi^\text{v}_f$}
	\newcommand{{\vhd}}{$\chi^\text{v}_d$}
	\newcommand{{\vpd}}{$\Delta^\text{v}_d$}
	\newcommand{{\ved}}{$\epsilon^\text{v}_d$}
	\newcommand{{\vved}}{$\varepsilon^\text{v}_d$}
	\newcommand{{\tr}}{{\rm tr}}
	\newcommand{\pprl}{Phys. Rev. Lett. \ }
	\newcommand{\pprb}{Phys. Rev. {B}}

\title {Topological Degeneracy Induced Flat Bands in two-Dimensional Holed Systems}
\author{Yuge Chen}
\thanks{These three authors contributed equally}
\affiliation{Institute for Quantum Science and Technology, Shanghai University, Shanghai 200444, China}
\affiliation{Beijing National Laboratory for Condensed Matter Physics and Institute of Physics,
	Chinese Academy of Sciences, Beijing 100190, China}

\author{Hui Yu}
\thanks{These three authors contributed equally}
\affiliation{Beijing National Laboratory for Condensed Matter Physics and Institute of Physics,
	Chinese Academy of Sciences, Beijing 100190, China}
	
\author{Yun-Peng Huang}
\thanks{These three authors contributed equally}
\affiliation{Beijing National Laboratory for Condensed Matter Physics and Institute of Physics,
	Chinese Academy of Sciences, Beijing 100190, China}
 
\author{Zhen-Yu Zheng}
\affiliation{Beijing National Laboratory for Condensed Matter Physics and Institute of Physics,
	Chinese Academy of Sciences, Beijing 100190, China}

\author{Jiangping Hu}
\email{jphu@iphy.ac.cn}
\affiliation{Beijing National Laboratory for Condensed Matter Physics and Institute of Physics,
	Chinese Academy of Sciences, Beijing 100190, China}
\affiliation{Kavli Institute of Theoretical Sciences, University of Chinese Academy of Sciences,
	Beijing, 100190, China}
	 \affiliation{New Cornerstone Science Laboratory, 
	Beijing, 100190, China}

\date{\today}

\begin{abstract}
Systems hosting flat bands offer a powerful platform for exploring strong correlation physics. Theoretically  topological degeneracy rising in systems with non-trivial topological orders on periodic manifolds of non-zero genus can generate ideal flat bands. However, experimental realization of such geometrically engineered systems is very difficult. In this work, we demonstrate that flat planes with strategically patterned hole defects can engineer ideal flat bands. We constructing two families of models, singular flat band systems where degeneracy is stabilized by non-contractible loop excitations tied to hole defects and perfectly nested van Hove systems where degeneracy arises from line excitations in momentum space. These models circumvent the need for exotic manifolds while retaining the essential features of topological flat bands. By directly linking defect engineering to degeneracy mechanisms, our results establish a scalable framework for experimentally accessible flat band design.
\end{abstract}
%\pacs{}
\maketitle

%\textit{Introduction}
\textit{Introduction.---}
Topological order \cite{wenrmp,kosterlitz2018ordering,thouless1982quantized,kitaev2003fault,kitaev2006anyons,levin2005string,moore1991nonabelions,nayak2008non,haldane1988model,hasan2010colloquium,chen2010local,senthil2000z,fidkowski2011topological,wen1990topological,read2000paired,kitaev2001unpaired,schnyder2008classification,fu2007topological} is a profound concept in condensed matter physics, describing phases of matter that cannot be characterized by traditional symmetry-breaking mechanisms \cite{ginzburg2009theory}. Topological degeneracy is one of the most fundamental aspects of systems exhibiting topological order.
Topological degeneracy refers to the ground state degeneracy of a system with topological order, which grows exponentially with the genus of the manifold on which the system is defined \cite{TaoWu,wenniu,wenspindege}. 

The $Z_2$ topological order, also known as the toric code or surface code \cite{toriccode1,toriccode2,toriccode3}, has been extensively studied as a promising candidate for topological quantum computing \cite{kitaev2003fault,kitaev2006anyons,nayak2008non,sarma2015majorana,dennis2002topological}. On a torus, the ground state of the toric code exhibits two zero modes, each corresponding to the generators of the torus's fundamental group, as illustrated in Fig.~\ref{introG} (a). When the toric code is placed on a periodic genera surface, as demonstrated in Fig.~\ref{introG}(b), its zero modes corresponds to the generators in the fundamental group of the surface. If we treat each genus as a supercell, two degenerate flat bands will be formed by these zero modes. In Figs.~\ref{introG}(c) and (d), we further divide the surface, with each resulting region becoming a 2-dimensional plane with holes. The relationship between the zero modes and the boundaries of these holes has been extensively studied both theoretically and experimentally for the toric code \cite{Freedmansurfacecode,surfacepra,surfacecode1,surfacecode2}. 
%Additionally, the connection between degeneracy and boundary conditions in generic topologically ordered systems has also been explored \cite{topoorderbound}.

\begin{figure}
\centering
\includegraphics[width=8.5cm]{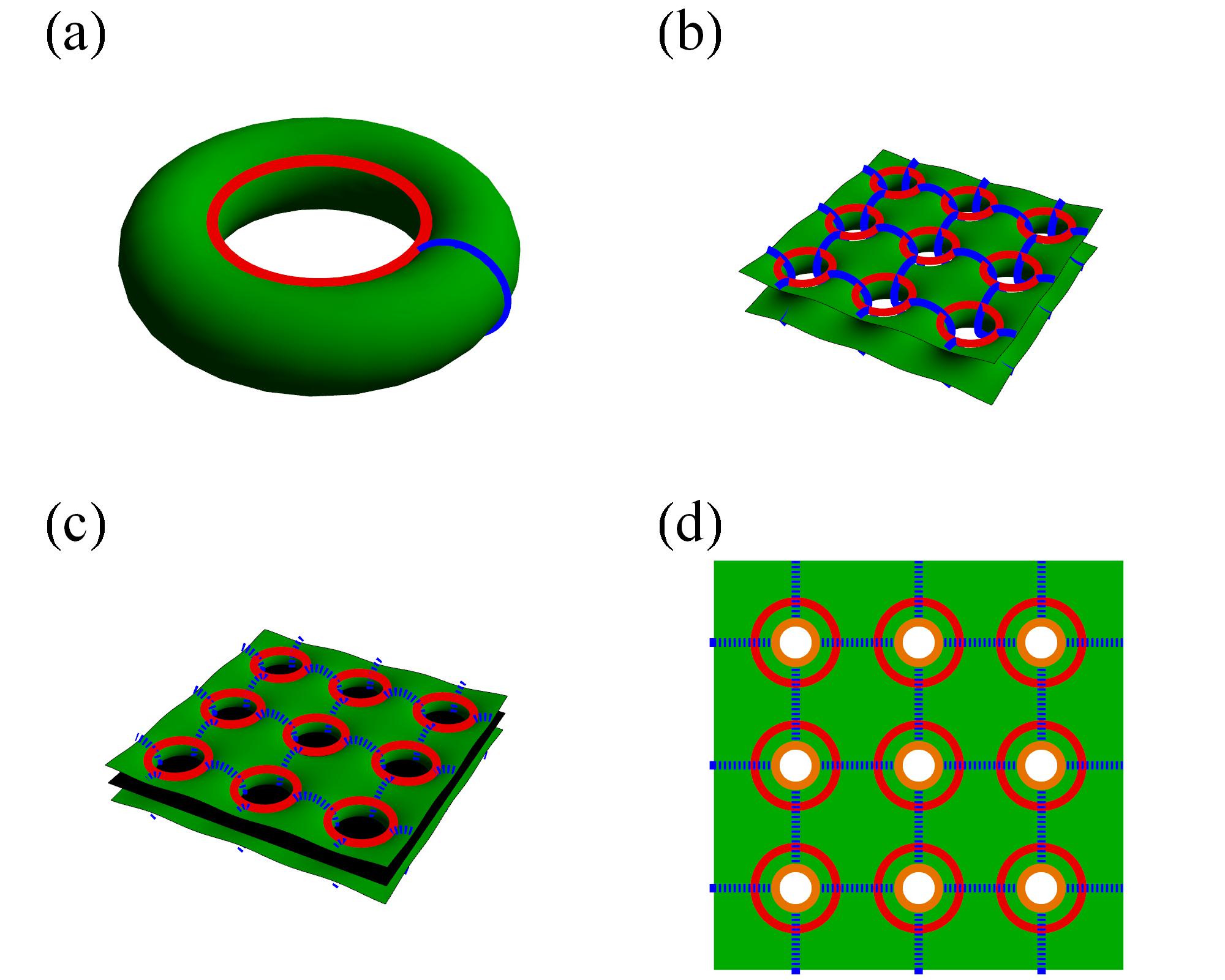}\\
\caption{(a) A torus surface, represented as a two-dimensional plane with periodic boundary conditions. The red and blue loops denote the two generators of its fundamental group, corresponding to two independent physical excitations. (b) A manifold with periodic genus. Due to the presence of $C_{4}$ symmetry, the number of blue loops has been doubled for illustration purposes. (c) The surface is separated into two layers by the black plane in the middle for experimental purposes. (d) A top-down view of one layer of the surface from (c). The white discs represent the holes. The orange circles indicate the boundary of these holes.}
\label{introG}
\end{figure}

In contrast to the toric code, systems with more complex topological order typically exhibit non-Abelian exchange phases \cite{kitaev2003fault,nayak2008non}, indicating that their ground states are strongly-correlated and cannot be described within a quasi-particle framework. This strong correlation makes the construction of such topologically ordered systems on a lattice particularly challenging. However, in this work, we show that flat bands can emerge in systems exhibiting topological degeneracy without necessarily requiring non-Abelian exchange phases. 

Before identifying potential candidates for such a system, we first provide a brief introduction to non-twisted flat band systems. To avoid any ambiguity, we will use the term "excitation" to refer to all  particle states of the system throughout the paper. In a non-twisted system, the presence of compact localized excitations (CLEs)—where the wave function is confined to a finite spatial region—necessarily implies the existence of a flat band that fully spans the Brillouin zone  \cite{Sutherlandcls,Flachcls1,Readcls1,Readcls2,Andocls,liuxin,cyg,wu08,SFBprb,SFBAPX,frederic1,frederic2,frederic3,frederic4,Julien1,Julien2,Julien3,Julien4,Dias1,Dias2,referee1-1,referee1-2,referee1-3,referee1-4,referee1-5,referee1-6-2,referee1-7-2,CLS0,CLS2,he1,he2,LiuWu,holed1,YYFBliu,liufeng1,liufeng2,Lieb,LiuLieb}. On the other hand, the existence of a flat band in the system guarantees the presence of CLEs \cite{SFBprb}. However, CLEs do not always form a complete basis for the flat band. In such cases, the flat band’s wave function exhibits a singular point, accompanied by two non-contractible loop excitations (NLEs) that are linearly independent of the CLEs. Such systems are referred to as singular flat band systems \cite{wu08,SFBprb,SFBAPX}. The NLEs correspond to non-contractible loops when singular flat band systems are placed on a torus. In analogy to topologically ordered systems, these systems display robust boundary modes around holes and line modes connecting boundaries in holed geometries, as evident in Fig.~\ref{holedlieb}(c) and (d). Previous experiments have successfully observed the presence of both types of modes in various singular flat band systems \cite{linestate0,linestate1,linestate2,opticalSFB1,opticalSFB2,opticalSFB3}.

In singular flat band systems, NLEs are inherently degenerate with the flat bands formed by CLEs.  Motivated by this fact, we investigate an alternative class of two-dimensional systems where excitations are neither protected by a flat band nor by topological order, yet exhibit similar degeneracy. Given that non-contractible loops are one-dimensional objects, the excitations corresponding to these loops must form a degenerate subspace along the orthogonal direction. This implies that the Fermi surface must include a straight line extending across the entire Brillouin zone. This straight line and its counterparts generated by the crystalline symmetry of the system, intersect at specific points in momentum space. These intersection points are typically identified as van Hove singularities. Consequently, a perfect nesting van Hove system can also achieve flat bands through a similar mechanism. 

In this Letter, we establish that embedding the Lieb lattice into a topologically non-trivial manifold creates a direct correspondence between non-contractible loop excitations and the generators of the fundamental group. This relationship reveals that singular flat band systems inherently exhibit topological degeneracy. Furthermore, we demonstrate that the degeneracy of flat bands in the Lieb lattice is determined by the specific hole configurations. Using the square lattice—a paradigmatic example of a perfectly nested van Hove system—we identify the boundary conditions necessary for the emergence of line modes. Building on this insight, we design a square lattice system with strategically placed holes and compute its energy spectrum. Our numerical results confirm the predicted appearance of zero-energy flat bands, which arise from line modes connecting pairs of holes.

\textit{Lieb lattice on a 1-genus surface.---}
The Hamiltonian of the Lieb lattice\cite{Lieb,LiuLieb,Tsai_2015,FCILieb}, considering only nearest-neighbor hopping, is given by:
\begin{equation}\label{Hlieb}
H=t\sum_{<i,j>}\left(c_i^\dagger c_j + h.c. \right)+\mu\sum_{\substack{i\ \in \\ corner}} c_{i}^\dagger c_{i}
\end{equation}
Here $c^{\dagger}_{i}$ and $c_{i}$ represent the electron creation and annihilation operators at the $i$-th lattice site. $t$ is the hopping strength and $\mu$ is the chemical potential defined on the corners of each plaquette.  A schematic diagram of the Lieb lattice with a CLE and NLE is shown in Fig.~\ref{Lieb}(a). When $\mu \neq 0$, the triple degeneracy at the M point splits into two parts: a single dispersive band and a band touching point where the flat band intersects with another dispersive band, as illustrated in Fig.~\ref{Lieb}(b).

\begin{figure}
\centering
  \includegraphics[width=8.5cm]{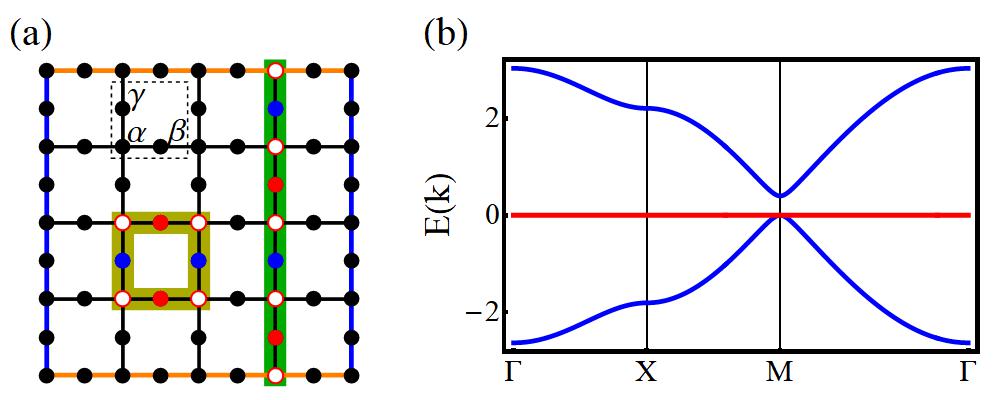}\\
  \caption{(a) A CLE (dark yellow) and NLE (green) in a Lieb lattice. The red circles mark the wave-function zeros for both states, while the red and blue dots represent the $+$ and $-$ phases of each state, respectively. The region enclosed by the dashed line denotes a unit cell. Boundaries of the same color are associated with one another. (b) The energy dispersion of the Lieb lattice with a chemical potential $\mu=0.4t$. The flat band is highlighted by the red line.}\label{Lieb}
\end{figure}

The wave-function of the flat band can be written as:
\begin{equation}\label{psi_wf}
\psi^\dagger(\vec k)=\frac{1}{Z}\left(cos(k_{y}a)c^\dagger_{\beta}(\vec k)-cos(k_{x}a)c^\dagger_{\gamma}(\vec k)\right)
\end{equation}
Where $\vec k = (k_{x},k_{y})$, $Z$ is the normalization constant, and $a$ denotes the lattice spacing. The electron creation operator with momentum $\vec k$ can be expanded as $c^\dagger_{i}(\vec k)=\frac{1}{\sqrt{N}}\sum_r e^{i\vec k\vec r}c^\dagger_{i}(\vec r)$, where $i$ runs over $\alpha,\beta,\gamma$, labeled within a unit cell of Lieb lattice, as depicted in Fig.~\ref{Lieb}(a), and $N$ is total number of unit cells. One can then verify that the flat band wave function at M point is zero. Furthermore, since the limits of the wave function as it approaches the M point from all directions form a 2-dimensional Hilbert space \cite{SFBprb,SFBAPX}, the M point is a singularity of the flat band's wave function. Thus, the Lieb lattice is a singular flat band system with a double degeneracy at the M point, provided the flatness is not broken. Since the momentum at the singular point of the Lieb lattice is non-zero, the NLEs can only arise when the horizontal and vertical directions of the lattice have even periodicities. This differs from the Kagome and Dice lattices, where the singular points occur at the $\Gamma$ point.

\textit{Lieb lattice on a 2-genus surface.---}
As introduced previously, the linearly independent NLEs of a Lieb lattice with periodic boundary conditions can be geometrically interpreted as non-contractible loops of the torus. As a result, when the Lieb lattice is placed on a 2-genus surface, there are four linearly independent NLEs, as seen in Fig.~\ref{Lieb3d}(b). To validate this, we diagonalize the Hamiltonian in Eq.~\ref{Hlieb} with $\mu \neq 0$ to obtain all the zero-energy states, which collectively form the Hilbert space $V_{E=0}$. By determining the Hilbert space of all CLEs, denoted as $V_{CLE}$, the number of linearly independent NLEs, $N_{NLE}$, can be expressed as:

\begin{figure}
\centering
  \includegraphics[width=8.5cm]{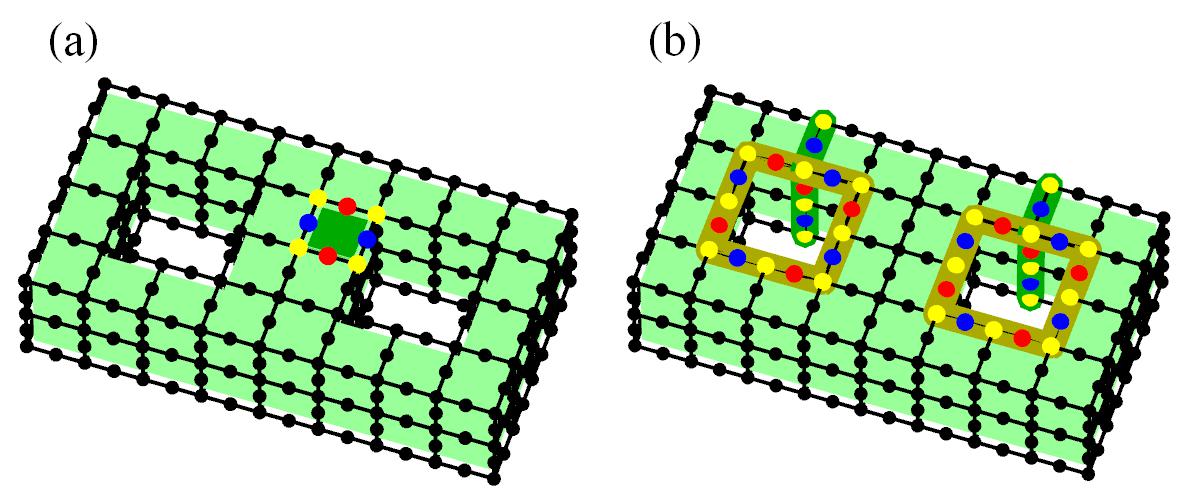}\\
  \caption{(a) Lieb lattice tiling on a box with two holes. The area highlighted in darker green represents an example of the CLE in the system. (b) Four linear independent NLEs emerge in the system. The yellow dots mark the wave-function zeros for CLEs and NLEs, while the red and blue dots correspond to the $\pm$ phases of each state, respectively.}\label{Lieb3d}
\end{figure}

\begin{equation}
N_{NLE}=dim((V_{E=0}^C\oplus V_{CLE})^C)
\end{equation}\label{nnle}
Here $V^C$ represents the complement of the space $V$ within the entire Hilbert space, which encompasses all possible states of the Lieb lattice. $dim$ denotes the dimension of the space. Consider a CLE wave function for a system with a singular flat band on a torus at position $\vec r$, denoted as $\psi_{CLE}(\vec r)$, its Fourier transformation always produce a finite-order Laurent polynomial $g(k)$ of $e^{ika}$ since the CLE is confined to a finite spatial region. Here, $g(k)$ is an eigenstate of the momentum operator with momentum $k$, and thus it must include the momentum wave function $\psi(k)$ from Eq.~\ref{psi_wf} as a factor. Since the $\psi(k)$ is also a Laurent polynomial of $e^{ika}$, we can express $g(k)$ as $f(k)\psi(k)$, where $f(k)$ is also a finite-order Laurent polynomial of $e^{ika}$. Consequently, the wave function $\psi_{CLE}(r)$ can be described as:
\begin{equation}\label{CLE_expand}
\psi_{CLE}(r)=\hat F^{-1}(f(k)\psi(k))=\sum _i c_i\phi \left(\vec r+r_i \vec a\right) 
\end{equation}
where $\{c_i\}$ and $\{r_i\}$ are the coefficients and degrees of each term in the Laurent polynomial $f(k)$, and $\phi(\vec r)$ represents the CLE localized on a single plaquette at $\vec r$, as highlighted in Fig.~\ref{Lieb3d}(a). $\hat F^{-1}$ denotes the inverse Fourier transform. It follows that $\psi_{CLE}(\vec r)$ is always a linear combination of finite $\phi(\vec r)$-terms, each centered at different plaquettes.

Since the local structures of surfaces with different genera are identical, any CLE defined on a 2-genus surface can also be expressed in the form given by Eq.~\ref{CLE_expand}. In particular, the Hilbert space of such a CLE is spanned by $\phi(\vec r)$s evaluated at all plaquettes. Next we compute all terms on the right hand side of Eq.~\ref{nnle} and find $N_{NLE}=4$. To further verify NLEs illustrated in Fig.~\ref{Lieb3d}(b) are linearly independent, we examine the following equation:
\begin{equation}
dim(V_0)=N_{NLE}-dim((V_{E=0}^C\oplus V_{CLE}\oplus V_0)^C),
\end{equation}
where $V_0$ is the Hilbert space spanned by different combinations of NLEs. This relationship confirms that the four NLEs are indeed linearly independent.

\begin{figure}
\centering
 \includegraphics[width=8.5cm]{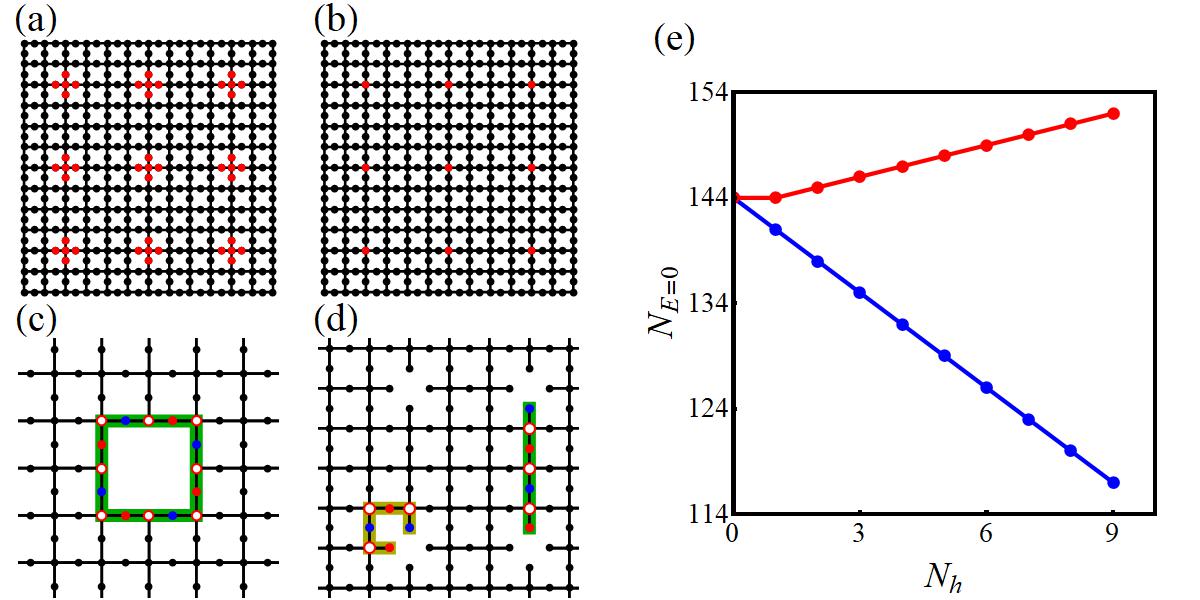}\\
 \caption{(a) An open-boundary Lieb lattice containing nine potential holes. Each interconnected set of red lattices sites corresponds to a potential hole. (b) An open-boundary Lieb lattice featuring nine potential holes, where each red lattice site represents a potential hole. (c) An example of NLE (darker green) around a hole in a Lieb lattice, corresponding to the configuration shown in panel (a). (d) An example of CLE (darker yellow) and NLE (darker green) emerging in a Lieb lattice, corresponding to the configuration illustrated in panel (b). The red circles mark the wave-function zeros for CLEs and NLEs, while the red and blue dots correspond to the $+$ and $-$ phases of each state, respectively. (e) The relationship between the number of all types zero-energy excitations, $N_{E=0}$, and the number of holes $N_{h}$ in two defect configurations. The blue line indicate configuration (a), while the red line corresponds to configuration (b). }\label{holedlieb}
\end{figure}

\textit{Lieb lattice with defects.---}
The topological degeneracy of the Lieb lattice can also be manifested in an open-boundary Lieb lattice with multiple hole defects. Similar to the toric code \cite{surfacepra}, the Lieb lattice can be punctured in two physically distinct ways, leading to two different types of boundary conditions. These boundary conditions are crucial in shaping the spatial configurations of CLE and NLE. As illustrated in Fig.~\ref{holedlieb}(a) and (c), removing five lattice sites to create a hole causes four CLEs to vanish while giving rise to an NLE (robust boundary mode) localized around the hole. On the other hand, Fig.~\ref{holedlieb}(b) and (d) demonstrate that introducing a hole defect by removing a single lattice site gives rise to new NLEs without altering the number of CLEs.  Notably, in this case, the NLE (line mode) form connections between two holes rather than being confined to a single hole. To further explore the relationship between topological degeneracy and the number of hole defects, we calculate the number of zero-energy states, $N_{E=0}$, for varying numbers of holes, $N_{h}$, under the two distinct puncturing schemes. Using the Hamiltonian Eq.~\ref{Hlieb} with $\mu=0.1t$, we find that $N_{E=0}$ exhibits a linear dependence on $N_{h}$. Specifically, $N_{E=0}$ decreases as $N_{h}$ increases for the defect configuration shown in Fig.~\ref{holedlieb}(a), while it grows with $N_{h}$ for the configuration depicted in Fig.~\ref{holedlieb}(b). We also note that the number of zero-energy states remains unchanged when only a single hole defect present in the system for the configuration in Fig.~\ref{holedlieb}(b). This is because the NLEs require two holes to form, as a single boundary cannot support linearly independent line excitations. In contrast, for boundary modes, the zero-energy degeneracy remains linearly dependent on the number of holes, even for a single hole. A similar phenomenon has recently been reported in the kagome lattice \cite{opticalSFB3}, where the flat-band degeneracy equals the sum of compact localized states and inner-robust boundary modes. Unlike the Lieb lattice, the kagome system lacks protected line modes due to the absence of boundary decoration. Our results establish a universal counting rule for singular flat-band degeneracy in holed systems:
\begin{equation}
N_{FBE}=N_{CLE}+N_{RBM}+N_{LM}
\end{equation}
where $N_{FBE}$ counts all flat-band excitations, $N_{CLE}$ is the number of compact localized excitations, while $N_{RBM}$ and $N_{LM}$ account for the linearly independent robust boundary modes and line modes, respectively.

\textit{Perfect nesting van Hove systems.---}
Flat band structures can also arise from perfect nesting van Hove systems through a mechanism analogous to the one described earlier. The half-filled square lattice with nearest-neighbor hopping represents the simplest example of such a system, exhibiting perfect nesting and providing an ideal platform for engineering flat bands. The Fermi surface of the system comprises four straight lines connecting the four K points, which coincide with the van Hove points, as shown in Fig.~\ref{sqstate}(a). When periodic boundary conditions are applied in both the vertical and horizontal directions, the square lattice system supports loop excitation as illustrated in Fig.~\ref{sqstate}(b). However, if the vertical direction is treated as an open boundary while periodicity is preserved in the horizontal direction, two distinct scenarios emerge depending on the alignment of the top and bottom boundaries. In the matched boundary case, the number of layers between the top and bottom boundaries is odd, resulting in a configuration where the boundaries align consistently. On the other hand, the mismatched boundary case arises when the number of layers between the top and bottom boundaries is even, leading to a misalignment of the boundaries. These two configurations are depicted in Fig.~\ref{sqstate}(c) and (d), respectively. In the case of matched boundaries, line excitations connecting the top and bottom boundaries emerge as eigenstates of the Hamiltonian. In contrast, when the boundaries are mismatched, the wave function of the line excitations cannot undergo destructive interference at the top green points in Fig.~\ref{sqstate}(d). As a result, these excitations are no longer eigenstates of the system.

\begin{figure}
\centering
  \includegraphics[width=8.5cm]{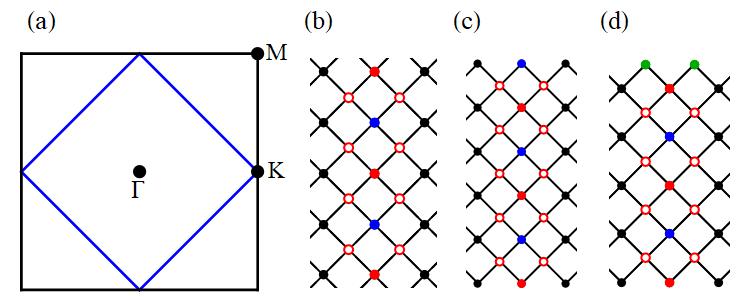}\\
  \caption{(a) The Brillouin zone of a square lattice. The blue line represents the Fermi surface of the square lattice at half-filling, corresponding to the energy level $E=0$. (b) The spatial configuration of a zero-energy loop excitation on a square lattice with periodic boundary conditions applied in both directions. Panels (c) and (d) illustrate the spatial configurations of line excitations under matched and unmatched boundary condition in the vertical direction, respectively, while periodicity is maintained in the horizontal direction. The color labeling follows the same convention as in Fig.~\ref{Lieb}(a).}\label{sqstate}
\end{figure}

Since line excitations can emerge as eigenstates of the system when the top and bottom boundaries are matched, we now leverage this property to construct a flat band by introducing edge-decorated holes into the square lattice.  A schematic diagram of such a holed square lattice is provided in Fig.~\ref{sqexample}(a). When we calculate the energy spectrum of this system, we observe that three bands completely overlap at $E=0$, corresponding to the van Hove energy of the square lattice. This triple degeneracy arises because four distinct diagonal lines can be drawn outward from each hole defect, with each line shared between two adjacent holes. This sharing results in two independent line excitations per hole, contributing two of the flat bands at zero energy. The third flat band, however, depends on the separation between neighboring holes and is not intrinsic, as its existence is sensitive to the specific geometric arrangement of the holes.

As seen in Fig.~\ref{sqexample}(b), the upper and lower dispersive bands form Dirac cones at specific points in momentum space. These Dirac cones disappear when the hopping strength $t$ becomes direction-dependent ($t_{x}\neq t_{y}$). Consequently, the middle flat bands become fully isolated within the resulting energy gap. In this case, the system no longer functions as a perfect nesting van Hove system, and the one-dimensional excitations characteristic cease to exist. Although flat bands may still persist in such a system, this aspect falls outside the scope of our current discussion.  

\begin{figure}
\centering
  \includegraphics[width=8.5cm]{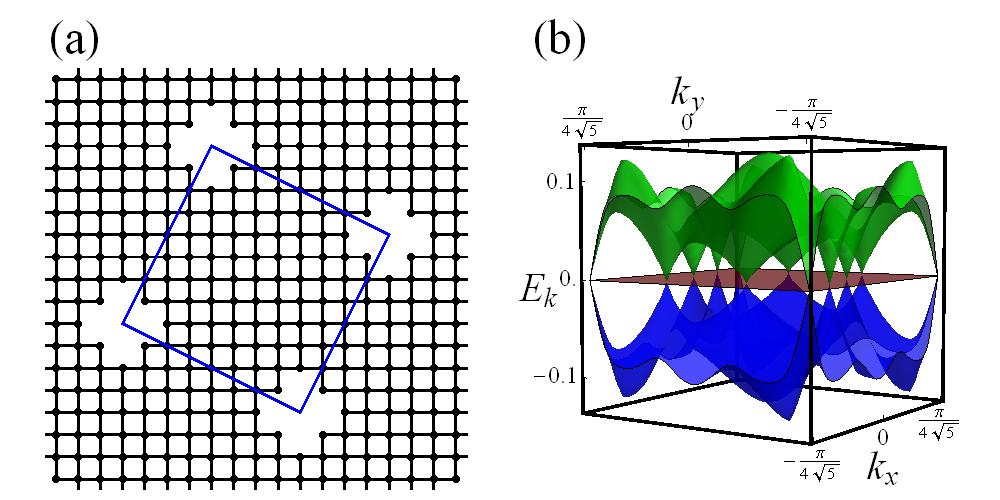}\\
  \caption{(a) A schematic drawing of a holed square lattice. The region enclosed by the blue lines represents the super-cell of the holed square lattice. (b) Energy spectrum of the five bands closest to zero energy, with the flat bands at $E=0$ being triple-degenerate.}\label{sqexample}
\end{figure}

\textit{Conclusion and discussion.---}
In this Letter, we demonstrate that placing the Lieb lattice on a topologically non-trivial manifold directly connects non-contractible loop excitations to the generators of the fundamental group, proving that singular flat band systems inherently host topological degeneracy. The flat band degeneracy depends critically on the lattice’s hole configurations. While this conclusion aligns with recent findings in Kagome systems \cite{opticalSFB3}, we have refined the specific formulas governing flat band degeneracy. Strikingly, quasi-1D Lieb rings exhibit entanglement entropy scaling governed by a free complex fermion conformal field theory, with sharp jumps at van Hove singularities as the Fermi surface varies (see Supplementary Material). We further derive the exact boundary conditions required for line modes in a square lattice—a paradigmatic example of a perfectly nested van Hove system. Inspired by this result, we engineer a square lattice with strategically placed holes, yielding a spectrum with three zero-energy flat bands. Two of these bands arise from independent line excitations and persist even when anisotropy breaks perfect nesting. Recently, several articles \cite{holed1,YYFBliu,liufeng1,liufeng2} have explored flat bands in other 2-dimensional systems with periodic holes. However, their approaches primarily involve replacing lattice sites with atomic substitutes, which differs significantly from our framework. In summary, these systems not only deepen our understanding of flat band formation in generic physical systems but also provide a versatile platform for engineering flat bands through precise geometric design.

\textit{Acknowledgement.---}
This work is supported by the Ministry of Science and Technology  (Grant No. 2022YFA1403901), the National Natural Science Foundation of China (Grant No. NSFC-12494594, No. NSFC-11888101, No. NSFC-12174428), the Strategic Priority Research Program of the Chinese Academy of Sciences (Grant No. XDB28000000), the New Cornerstone Investigator Program, the Chinese Academy of Sciences through the Youth Innovation Promotion Association (Grant No. 2022YSBR-048) and the Shanghai Science and Technology Innovation Action Plan (Grant No. 24LZ1400800). We want to thank Prof Qian Niu, Yong-shi Wu, Yue Yu and Dr. Yuhao Gu for helpful discussion. 

\textit{Data availability.}  Numerical data for this manuscript are publicly accessible in Ref. \cite{data-availbale}.

\bibliographystyle{apsrev4-2}
\bibliography{reference}

\begin{thebibliography}{81}%
\makeatletter
\providecommand \@ifxundefined [1]{%
 \@ifx{#1\undefined}
}%
\providecommand \@ifnum [1]{%
 \ifnum #1\expandafter \@firstoftwo
 \else \expandafter \@secondoftwo
 \fi
}%
\providecommand \@ifx [1]{%
 \ifx #1\expandafter \@firstoftwo
 \else \expandafter \@secondoftwo
 \fi
}%
\providecommand \natexlab [1]{#1}%
\providecommand \enquote  [1]{``#1''}%
\providecommand \bibnamefont  [1]{#1}%
\providecommand \bibfnamefont [1]{#1}%
\providecommand \citenamefont [1]{#1}%
\providecommand \href@noop [0]{\@secondoftwo}%
\providecommand \href [0]{\begingroup \@sanitize@url \@href}%
\providecommand \@href[1]{\@@startlink{#1}\@@href}%
\providecommand \@@href[1]{\endgroup#1\@@endlink}%
\providecommand \@sanitize@url [0]{\catcode `\\12\catcode `\$12\catcode `\&12\catcode `\#12\catcode `\^12\catcode `\_12\catcode `\%12\relax}%
\providecommand \@@startlink[1]{}%
\providecommand \@@endlink[0]{}%
\providecommand \url  [0]{\begingroup\@sanitize@url \@url }%
\providecommand \@url [1]{\endgroup\@href {#1}{\urlprefix }}%
\providecommand \urlprefix  [0]{URL }%
\providecommand \Eprint [0]{\href }%
\providecommand \doibase [0]{https://doi.org/}%
\providecommand \selectlanguage [0]{\@gobble}%
\providecommand \bibinfo  [0]{\@secondoftwo}%
\providecommand \bibfield  [0]{\@secondoftwo}%
\providecommand \translation [1]{[#1]}%
\providecommand \BibitemOpen [0]{}%
\providecommand \bibitemStop [0]{}%
\providecommand \bibitemNoStop [0]{.\EOS\space}%
\providecommand \EOS [0]{\spacefactor3000\relax}%
\providecommand \BibitemShut  [1]{\csname bibitem#1\endcsname}%
\let\auto@bib@innerbib\@empty
%</preamble>
\bibitem [{\citenamefont {Wen}(2017)}]{wenrmp}%
  \BibitemOpen
  \bibfield  {author} {\bibinfo {author} {\bibfnamefont {X.-G.}\ \bibnamefont {Wen}},\ }\href {https://doi.org/10.1103/RevModPhys.89.041004} {\bibfield  {journal} {\bibinfo  {journal} {Rev. Mod. Phys.}\ }\textbf {\bibinfo {volume} {89}},\ \bibinfo {pages} {041004} (\bibinfo {year} {2017})}\BibitemShut {NoStop}%
\bibitem [{\citenamefont {Kosterlitz}\ and\ \citenamefont {Thouless}(2018)}]{kosterlitz2018ordering}%
  \BibitemOpen
  \bibfield  {author} {\bibinfo {author} {\bibfnamefont {J.~M.}\ \bibnamefont {Kosterlitz}}\ and\ \bibinfo {author} {\bibfnamefont {D.~J.}\ \bibnamefont {Thouless}},\ }in\ \href@noop {} {\emph {\bibinfo {booktitle} {Basic Notions Of Condensed Matter Physics}}}\ (\bibinfo  {publisher} {CRC Press},\ \bibinfo {year} {2018})\ pp.\ \bibinfo {pages} {493--515}\BibitemShut {NoStop}%
\bibitem [{\citenamefont {Thouless}\ \emph {et~al.}(1982)\citenamefont {Thouless}, \citenamefont {Kohmoto}, \citenamefont {Nightingale},\ and\ \citenamefont {den Nijs}}]{thouless1982quantized}%
  \BibitemOpen
  \bibfield  {author} {\bibinfo {author} {\bibfnamefont {D.~J.}\ \bibnamefont {Thouless}}, \bibinfo {author} {\bibfnamefont {M.}~\bibnamefont {Kohmoto}}, \bibinfo {author} {\bibfnamefont {M.~P.}\ \bibnamefont {Nightingale}},\ and\ \bibinfo {author} {\bibfnamefont {M.}~\bibnamefont {den Nijs}},\ }\href {https://doi.org/10.1103/PhysRevLett.49.405} {\bibfield  {journal} {\bibinfo  {journal} {Phys. Rev. Lett.}\ }\textbf {\bibinfo {volume} {49}},\ \bibinfo {pages} {405} (\bibinfo {year} {1982})}\BibitemShut {NoStop}%
\bibitem [{\citenamefont {Kitaev}(2003)}]{kitaev2003fault}%
  \BibitemOpen
  \bibfield  {author} {\bibinfo {author} {\bibfnamefont {A.}~\bibnamefont {Kitaev}},\ }\href {https://doi.org/https://doi.org/10.1016/S0003-4916(02)00018-0} {\bibfield  {journal} {\bibinfo  {journal} {Annals of Physics}\ }\textbf {\bibinfo {volume} {303}},\ \bibinfo {pages} {2} (\bibinfo {year} {2003})}\BibitemShut {NoStop}%
\bibitem [{\citenamefont {Kitaev}(2006)}]{kitaev2006anyons}%
  \BibitemOpen
  \bibfield  {author} {\bibinfo {author} {\bibfnamefont {A.}~\bibnamefont {Kitaev}},\ }\href {https://doi.org/https://doi.org/10.1016/j.aop.2005.10.005} {\bibfield  {journal} {\bibinfo  {journal} {Annals of Physics}\ }\textbf {\bibinfo {volume} {321}},\ \bibinfo {pages} {2} (\bibinfo {year} {2006})},\ \bibinfo {note} {january Special Issue}\BibitemShut {NoStop}%
\bibitem [{\citenamefont {Levin}\ and\ \citenamefont {Wen}(2005)}]{levin2005string}%
  \BibitemOpen
  \bibfield  {author} {\bibinfo {author} {\bibfnamefont {M.~A.}\ \bibnamefont {Levin}}\ and\ \bibinfo {author} {\bibfnamefont {X.-G.}\ \bibnamefont {Wen}},\ }\href {https://doi.org/10.1103/PhysRevB.71.045110} {\bibfield  {journal} {\bibinfo  {journal} {Phys. Rev. B}\ }\textbf {\bibinfo {volume} {71}},\ \bibinfo {pages} {045110} (\bibinfo {year} {2005})}\BibitemShut {NoStop}%
\bibitem [{\citenamefont {Moore}\ and\ \citenamefont {Read}(1991)}]{moore1991nonabelions}%
  \BibitemOpen
  \bibfield  {author} {\bibinfo {author} {\bibfnamefont {G.}~\bibnamefont {Moore}}\ and\ \bibinfo {author} {\bibfnamefont {N.}~\bibnamefont {Read}},\ }\href {https://doi.org/10.1016/0550-3213(91)90407-O} {\bibfield  {journal} {\bibinfo  {journal} {Nuclear Phys. B}\ }\textbf {\bibinfo {volume} {360}},\ \bibinfo {pages} {362} (\bibinfo {year} {1991})}\BibitemShut {NoStop}%
\bibitem [{\citenamefont {Nayak}\ \emph {et~al.}(2007)\citenamefont {Nayak}, \citenamefont {Simon}, \citenamefont {Stern}, \citenamefont {Freedman},\ and\ \citenamefont {Das~Sarma}}]{nayak2008non}%
  \BibitemOpen
  \bibfield  {author} {\bibinfo {author} {\bibfnamefont {D.~C.}\ \bibnamefont {Nayak}}, \bibinfo {author} {\bibfnamefont {S.~H.}\ \bibnamefont {Simon}}, \bibinfo {author} {\bibfnamefont {A.}~\bibnamefont {Stern}}, \bibinfo {author} {\bibfnamefont {M.}~\bibnamefont {Freedman}},\ and\ \bibinfo {author} {\bibfnamefont {D.~S.}\ \bibnamefont {Das~Sarma}},\ }\href {https://www.microsoft.com/en-us/research/publication/non-abelian-anyons-topological-quantum-computation/} {\bibfield  {journal} {\bibinfo  {journal} {Strongly Correlated Electrons}\ }\textbf {\bibinfo {volume} {80}} (\bibinfo {year} {2007})}\BibitemShut {NoStop}%
\bibitem [{\citenamefont {Haldane}(1988)}]{haldane1988model}%
  \BibitemOpen
  \bibfield  {author} {\bibinfo {author} {\bibfnamefont {F.~D.~M.}\ \bibnamefont {Haldane}},\ }\href {https://doi.org/10.1103/PhysRevLett.61.2015} {\bibfield  {journal} {\bibinfo  {journal} {Phys. Rev. Lett.}\ }\textbf {\bibinfo {volume} {61}},\ \bibinfo {pages} {2015} (\bibinfo {year} {1988})}\BibitemShut {NoStop}%
\bibitem [{\citenamefont {Hasan}\ and\ \citenamefont {Kane}(2010)}]{hasan2010colloquium}%
  \BibitemOpen
  \bibfield  {author} {\bibinfo {author} {\bibfnamefont {M.~Z.}\ \bibnamefont {Hasan}}\ and\ \bibinfo {author} {\bibfnamefont {C.~L.}\ \bibnamefont {Kane}},\ }\href {https://doi.org/10.1103/RevModPhys.82.3045} {\bibfield  {journal} {\bibinfo  {journal} {Rev. Mod. Phys.}\ }\textbf {\bibinfo {volume} {82}},\ \bibinfo {pages} {3045} (\bibinfo {year} {2010})}\BibitemShut {NoStop}%
\bibitem [{\citenamefont {Chen}\ \emph {et~al.}(2010)\citenamefont {Chen}, \citenamefont {Gu},\ and\ \citenamefont {Wen}}]{chen2010local}%
  \BibitemOpen
  \bibfield  {author} {\bibinfo {author} {\bibfnamefont {X.}~\bibnamefont {Chen}}, \bibinfo {author} {\bibfnamefont {Z.-C.}\ \bibnamefont {Gu}},\ and\ \bibinfo {author} {\bibfnamefont {X.-G.}\ \bibnamefont {Wen}},\ }\href {https://doi.org/10.1103/PhysRevB.82.155138} {\bibfield  {journal} {\bibinfo  {journal} {Phys. Rev. B}\ }\textbf {\bibinfo {volume} {82}},\ \bibinfo {pages} {155138} (\bibinfo {year} {2010})}\BibitemShut {NoStop}%
\bibitem [{\citenamefont {Senthil}\ and\ \citenamefont {Fisher}(2000)}]{senthil2000z}%
  \BibitemOpen
  \bibfield  {author} {\bibinfo {author} {\bibfnamefont {T.}~\bibnamefont {Senthil}}\ and\ \bibinfo {author} {\bibfnamefont {M.~P.~A.}\ \bibnamefont {Fisher}},\ }\href {https://doi.org/10.1103/PhysRevB.62.7850} {\bibfield  {journal} {\bibinfo  {journal} {Phys. Rev. B}\ }\textbf {\bibinfo {volume} {62}},\ \bibinfo {pages} {7850} (\bibinfo {year} {2000})}\BibitemShut {NoStop}%
\bibitem [{\citenamefont {Fidkowski}\ and\ \citenamefont {Kitaev}(2011)}]{fidkowski2011topological}%
  \BibitemOpen
  \bibfield  {author} {\bibinfo {author} {\bibfnamefont {L.}~\bibnamefont {Fidkowski}}\ and\ \bibinfo {author} {\bibfnamefont {A.}~\bibnamefont {Kitaev}},\ }\href {https://doi.org/10.1103/PhysRevB.83.075103} {\bibfield  {journal} {\bibinfo  {journal} {Phys. Rev. B}\ }\textbf {\bibinfo {volume} {83}},\ \bibinfo {pages} {075103} (\bibinfo {year} {2011})}\BibitemShut {NoStop}%
\bibitem [{\citenamefont {Wen}(1990)}]{wen1990topological}%
  \BibitemOpen
  \bibfield  {author} {\bibinfo {author} {\bibfnamefont {X.~G.}\ \bibnamefont {Wen}},\ }\href {https://doi.org/10.1142/S0217979290000139} {\bibfield  {journal} {\bibinfo  {journal} {International Journal of Modern Physics B}\ }\textbf {\bibinfo {volume} {04}},\ \bibinfo {pages} {239} (\bibinfo {year} {1990})}\BibitemShut {NoStop}%
\bibitem [{\citenamefont {Read}\ and\ \citenamefont {Green}(2000)}]{read2000paired}%
  \BibitemOpen
  \bibfield  {author} {\bibinfo {author} {\bibfnamefont {N.}~\bibnamefont {Read}}\ and\ \bibinfo {author} {\bibfnamefont {D.}~\bibnamefont {Green}},\ }\href {https://doi.org/10.1103/PhysRevB.61.10267} {\bibfield  {journal} {\bibinfo  {journal} {Phys. Rev. B}\ }\textbf {\bibinfo {volume} {61}},\ \bibinfo {pages} {10267} (\bibinfo {year} {2000})}\BibitemShut {NoStop}%
\bibitem [{\citenamefont {Kitaev}(2001)}]{kitaev2001unpaired}%
  \BibitemOpen
  \bibfield  {author} {\bibinfo {author} {\bibfnamefont {A.~Y.}\ \bibnamefont {Kitaev}},\ }\href {https://doi.org/10.1070/1063-7869/44/10S/S29} {\bibfield  {journal} {\bibinfo  {journal} {Physics-Uspekhi}\ }\textbf {\bibinfo {volume} {44}},\ \bibinfo {pages} {131} (\bibinfo {year} {2001})}\BibitemShut {NoStop}%
\bibitem [{\citenamefont {Schnyder}\ \emph {et~al.}(2008)\citenamefont {Schnyder}, \citenamefont {Ryu}, \citenamefont {Furusaki},\ and\ \citenamefont {Ludwig}}]{schnyder2008classification}%
  \BibitemOpen
  \bibfield  {author} {\bibinfo {author} {\bibfnamefont {A.~P.}\ \bibnamefont {Schnyder}}, \bibinfo {author} {\bibfnamefont {S.}~\bibnamefont {Ryu}}, \bibinfo {author} {\bibfnamefont {A.}~\bibnamefont {Furusaki}},\ and\ \bibinfo {author} {\bibfnamefont {A.~W.~W.}\ \bibnamefont {Ludwig}},\ }\href {https://doi.org/10.1103/PhysRevB.78.195125} {\bibfield  {journal} {\bibinfo  {journal} {Phys. Rev. B}\ }\textbf {\bibinfo {volume} {78}},\ \bibinfo {pages} {195125} (\bibinfo {year} {2008})}\BibitemShut {NoStop}%
\bibitem [{\citenamefont {Fu}\ \emph {et~al.}(2007)\citenamefont {Fu}, \citenamefont {Kane},\ and\ \citenamefont {Mele}}]{fu2007topological}%
  \BibitemOpen
  \bibfield  {author} {\bibinfo {author} {\bibfnamefont {L.}~\bibnamefont {Fu}}, \bibinfo {author} {\bibfnamefont {C.~L.}\ \bibnamefont {Kane}},\ and\ \bibinfo {author} {\bibfnamefont {E.~J.}\ \bibnamefont {Mele}},\ }\href {https://doi.org/10.1103/PhysRevLett.98.106803} {\bibfield  {journal} {\bibinfo  {journal} {Phys. Rev. Lett.}\ }\textbf {\bibinfo {volume} {98}},\ \bibinfo {pages} {106803} (\bibinfo {year} {2007})}\BibitemShut {NoStop}%
\bibitem [{\citenamefont {Ginzburg}\ \emph {et~al.}(2009)\citenamefont {Ginzburg}, \citenamefont {Ginzburg},\ and\ \citenamefont {Landau}}]{ginzburg2009theory}%
  \BibitemOpen
  \bibfield  {author} {\bibinfo {author} {\bibfnamefont {V.~L.}\ \bibnamefont {Ginzburg}}, \bibinfo {author} {\bibfnamefont {V.~L.}\ \bibnamefont {Ginzburg}},\ and\ \bibinfo {author} {\bibfnamefont {L.}~\bibnamefont {Landau}},\ }\href@noop {} {\emph {\bibinfo {title} {On the theory of superconductivity}}}\ (\bibinfo  {publisher} {Springer},\ \bibinfo {year} {2009})\BibitemShut {NoStop}%
\bibitem [{\citenamefont {Tao}\ and\ \citenamefont {Wu}(1984)}]{TaoWu}%
  \BibitemOpen
  \bibfield  {author} {\bibinfo {author} {\bibfnamefont {R.}~\bibnamefont {Tao}}\ and\ \bibinfo {author} {\bibfnamefont {Y.-S.}\ \bibnamefont {Wu}},\ }\href {https://doi.org/10.1103/PhysRevB.30.1097} {\bibfield  {journal} {\bibinfo  {journal} {Phys. Rev. B}\ }\textbf {\bibinfo {volume} {30}},\ \bibinfo {pages} {1097} (\bibinfo {year} {1984})}\BibitemShut {NoStop}%
\bibitem [{\citenamefont {Wen}\ and\ \citenamefont {Niu}(1990)}]{wenniu}%
  \BibitemOpen
  \bibfield  {author} {\bibinfo {author} {\bibfnamefont {X.~G.}\ \bibnamefont {Wen}}\ and\ \bibinfo {author} {\bibfnamefont {Q.}~\bibnamefont {Niu}},\ }\href {https://doi.org/10.1103/PhysRevB.41.9377} {\bibfield  {journal} {\bibinfo  {journal} {Phys. Rev. B}\ }\textbf {\bibinfo {volume} {41}},\ \bibinfo {pages} {9377} (\bibinfo {year} {1990})}\BibitemShut {NoStop}%
\bibitem [{\citenamefont {Wen}(1989)}]{wenspindege}%
  \BibitemOpen
  \bibfield  {author} {\bibinfo {author} {\bibfnamefont {X.~G.}\ \bibnamefont {Wen}},\ }\href {https://doi.org/10.1103/PhysRevB.40.7387} {\bibfield  {journal} {\bibinfo  {journal} {Phys. Rev. B}\ }\textbf {\bibinfo {volume} {40}},\ \bibinfo {pages} {7387} (\bibinfo {year} {1989})}\BibitemShut {NoStop}%
\bibitem [{\citenamefont {Bravyi}\ and\ \citenamefont {Kitaev}(1998)}]{toriccode1}%
  \BibitemOpen
  \bibfield  {author} {\bibinfo {author} {\bibfnamefont {S.~B.}\ \bibnamefont {Bravyi}}\ and\ \bibinfo {author} {\bibfnamefont {A.~Y.}\ \bibnamefont {Kitaev}},\ }\href {https://arxiv.org/abs/quant-ph/9811052} {\bibinfo {title} {Quantum codes on a lattice with boundary}} (\bibinfo {year} {1998})\BibitemShut {NoStop}%
\bibitem [{\citenamefont {Beigi}\ \emph {et~al.}(2011)\citenamefont {Beigi}, \citenamefont {Shor},\ and\ \citenamefont {Whalen}}]{toriccode2}%
  \BibitemOpen
  \bibfield  {author} {\bibinfo {author} {\bibfnamefont {S.}~\bibnamefont {Beigi}}, \bibinfo {author} {\bibfnamefont {P.~W.}\ \bibnamefont {Shor}},\ and\ \bibinfo {author} {\bibfnamefont {D.}~\bibnamefont {Whalen}},\ }\href {https://doi.org/10.1007/s00220-011-1294-x} {\bibfield  {journal} {\bibinfo  {journal} {Communications in Mathematical Physics}\ }\textbf {\bibinfo {volume} {306}},\ \bibinfo {pages} {663} (\bibinfo {year} {2011})}\BibitemShut {NoStop}%
\bibitem [{\citenamefont {Kitaev}\ and\ \citenamefont {Kong}(2012)}]{toriccode3}%
  \BibitemOpen
  \bibfield  {author} {\bibinfo {author} {\bibfnamefont {A.}~\bibnamefont {Kitaev}}\ and\ \bibinfo {author} {\bibfnamefont {L.}~\bibnamefont {Kong}},\ }\href {https://doi.org/10.1007/s00220-012-1500-5} {\bibfield  {journal} {\bibinfo  {journal} {Communications in Mathematical Physics}\ }\textbf {\bibinfo {volume} {313}},\ \bibinfo {pages} {351} (\bibinfo {year} {2012})}\BibitemShut {NoStop}%
\bibitem [{\citenamefont {Sarma}\ \emph {et~al.}(2015)\citenamefont {Sarma}, \citenamefont {Freedman},\ and\ \citenamefont {Nayak}}]{sarma2015majorana}%
  \BibitemOpen
  \bibfield  {author} {\bibinfo {author} {\bibfnamefont {S.~D.}\ \bibnamefont {Sarma}}, \bibinfo {author} {\bibfnamefont {M.}~\bibnamefont {Freedman}},\ and\ \bibinfo {author} {\bibfnamefont {C.}~\bibnamefont {Nayak}},\ }\href {https://doi.org/10.1038/npjqi.2015.1} {\bibfield  {journal} {\bibinfo  {journal} {npj Quantum Information}\ }\textbf {\bibinfo {volume} {1}},\ \bibinfo {pages} {15001} (\bibinfo {year} {2015})}\BibitemShut {NoStop}%
\bibitem [{\citenamefont {Dennis}\ \emph {et~al.}(2002)\citenamefont {Dennis}, \citenamefont {Kitaev}, \citenamefont {Landahl},\ and\ \citenamefont {Preskill}}]{dennis2002topological}%
  \BibitemOpen
  \bibfield  {author} {\bibinfo {author} {\bibfnamefont {E.}~\bibnamefont {Dennis}}, \bibinfo {author} {\bibfnamefont {A.}~\bibnamefont {Kitaev}}, \bibinfo {author} {\bibfnamefont {A.}~\bibnamefont {Landahl}},\ and\ \bibinfo {author} {\bibfnamefont {J.}~\bibnamefont {Preskill}},\ }\href {https://doi.org/10.1063/1.1499754} {\bibfield  {journal} {\bibinfo  {journal} {Journal of Mathematical Physics}\ }\textbf {\bibinfo {volume} {43}},\ \bibinfo {pages} {4452} (\bibinfo {year} {2002})}\BibitemShut {NoStop}%
\bibitem [{\citenamefont {Freedman}(2001)}]{Freedmansurfacecode}%
  \BibitemOpen
  \bibfield  {author} {\bibinfo {author} {\bibfnamefont {M.~H.}\ \bibnamefont {Freedman}},\ }\href {https://doi.org/10.1007/s102080010006} {\bibfield  {journal} {\bibinfo  {journal} {Foundations of Computational Mathematics}\ }\textbf {\bibinfo {volume} {1}},\ \bibinfo {pages} {183} (\bibinfo {year} {2001})}\BibitemShut {NoStop}%
\bibitem [{\citenamefont {Fowler}\ \emph {et~al.}(2012)\citenamefont {Fowler}, \citenamefont {Mariantoni}, \citenamefont {Martinis},\ and\ \citenamefont {Cleland}}]{surfacepra}%
  \BibitemOpen
  \bibfield  {author} {\bibinfo {author} {\bibfnamefont {A.~G.}\ \bibnamefont {Fowler}}, \bibinfo {author} {\bibfnamefont {M.}~\bibnamefont {Mariantoni}}, \bibinfo {author} {\bibfnamefont {J.~M.}\ \bibnamefont {Martinis}},\ and\ \bibinfo {author} {\bibfnamefont {A.~N.}\ \bibnamefont {Cleland}},\ }\href {https://doi.org/10.1103/PhysRevA.86.032324} {\bibfield  {journal} {\bibinfo  {journal} {Phys. Rev. A}\ }\textbf {\bibinfo {volume} {86}},\ \bibinfo {pages} {032324} (\bibinfo {year} {2012})}\BibitemShut {NoStop}%
\bibitem [{\citenamefont {AI}(2023)}]{surfacecode1}%
  \BibitemOpen
  \bibfield  {author} {\bibinfo {author} {\bibfnamefont {G.~Q.}\ \bibnamefont {AI}},\ }\href {https://doi.org/10.1038/s41586-022-05434-1} {\bibfield  {journal} {\bibinfo  {journal} {Nature}\ }\textbf {\bibinfo {volume} {614}},\ \bibinfo {pages} {676} (\bibinfo {year} {2023})}\BibitemShut {NoStop}%
\bibitem [{\citenamefont {AI}(2021)}]{surfacecode2}%
  \BibitemOpen
  \bibfield  {author} {\bibinfo {author} {\bibfnamefont {G.~Q.}\ \bibnamefont {AI}},\ }\href {https://doi.org/10.1038/s41586-021-03588-y} {\bibfield  {journal} {\bibinfo  {journal} {Nature}\ }\textbf {\bibinfo {volume} {595}},\ \bibinfo {pages} {383} (\bibinfo {year} {2021})}\BibitemShut {NoStop}%
\bibitem [{\citenamefont {Sutherland}(1986)}]{Sutherlandcls}%
  \BibitemOpen
  \bibfield  {author} {\bibinfo {author} {\bibfnamefont {B.}~\bibnamefont {Sutherland}},\ }\href {https://doi.org/10.1103/PhysRevB.34.5208} {\bibfield  {journal} {\bibinfo  {journal} {Phys. Rev. B}\ }\textbf {\bibinfo {volume} {34}},\ \bibinfo {pages} {5208} (\bibinfo {year} {1986})}\BibitemShut {NoStop}%
\bibitem [{\citenamefont {Maimaiti}\ \emph {et~al.}(2017)\citenamefont {Maimaiti}, \citenamefont {Andreanov}, \citenamefont {Park}, \citenamefont {Gendelman},\ and\ \citenamefont {Flach}}]{Flachcls1}%
  \BibitemOpen
  \bibfield  {author} {\bibinfo {author} {\bibfnamefont {W.}~\bibnamefont {Maimaiti}}, \bibinfo {author} {\bibfnamefont {A.}~\bibnamefont {Andreanov}}, \bibinfo {author} {\bibfnamefont {H.~C.}\ \bibnamefont {Park}}, \bibinfo {author} {\bibfnamefont {O.}~\bibnamefont {Gendelman}},\ and\ \bibinfo {author} {\bibfnamefont {S.}~\bibnamefont {Flach}},\ }\href {https://doi.org/10.1103/PhysRevB.95.115135} {\bibfield  {journal} {\bibinfo  {journal} {Phys. Rev. B}\ }\textbf {\bibinfo {volume} {95}},\ \bibinfo {pages} {115135} (\bibinfo {year} {2017})}\BibitemShut {NoStop}%
\bibitem [{\citenamefont {Dubail}\ and\ \citenamefont {Read}(2015)}]{Readcls1}%
  \BibitemOpen
  \bibfield  {author} {\bibinfo {author} {\bibfnamefont {J.}~\bibnamefont {Dubail}}\ and\ \bibinfo {author} {\bibfnamefont {N.}~\bibnamefont {Read}},\ }\href {https://doi.org/10.1103/PhysRevB.92.205307} {\bibfield  {journal} {\bibinfo  {journal} {Phys. Rev. B}\ }\textbf {\bibinfo {volume} {92}},\ \bibinfo {pages} {205307} (\bibinfo {year} {2015})}\BibitemShut {NoStop}%
\bibitem [{\citenamefont {Read}(2017)}]{Readcls2}%
  \BibitemOpen
  \bibfield  {author} {\bibinfo {author} {\bibfnamefont {N.}~\bibnamefont {Read}},\ }\href {https://doi.org/10.1103/PhysRevB.95.115309} {\bibfield  {journal} {\bibinfo  {journal} {Phys. Rev. B}\ }\textbf {\bibinfo {volume} {95}},\ \bibinfo {pages} {115309} (\bibinfo {year} {2017})}\BibitemShut {NoStop}%
\bibitem [{\citenamefont {Aoki}\ \emph {et~al.}(1996)\citenamefont {Aoki}, \citenamefont {Ando},\ and\ \citenamefont {Matsumura}}]{Andocls}%
  \BibitemOpen
  \bibfield  {author} {\bibinfo {author} {\bibfnamefont {H.}~\bibnamefont {Aoki}}, \bibinfo {author} {\bibfnamefont {M.}~\bibnamefont {Ando}},\ and\ \bibinfo {author} {\bibfnamefont {H.}~\bibnamefont {Matsumura}},\ }\href {https://doi.org/10.1103/PhysRevB.54.R17296} {\bibfield  {journal} {\bibinfo  {journal} {Phys. Rev. B}\ }\textbf {\bibinfo {volume} {54}},\ \bibinfo {pages} {R17296} (\bibinfo {year} {1996})}\BibitemShut {NoStop}%
\bibitem [{\citenamefont {Liu}\ and\ \citenamefont {Liu}(2024)}]{liuxin}%
  \BibitemOpen
  \bibfield  {author} {\bibinfo {author} {\bibfnamefont {R.-H.}\ \bibnamefont {Liu}}\ and\ \bibinfo {author} {\bibfnamefont {X.}~\bibnamefont {Liu}},\ }\href {https://arxiv.org/abs/2412.15653} {\bibinfo {title} {Unified real-space construction scheme for flat bands based on symmetric compact localized states}} (\bibinfo {year} {2024}),\ \Eprint {https://arxiv.org/abs/2412.15653} {arXiv:2412.15653 [cond-mat.mes-hall]} \BibitemShut {NoStop}%
\bibitem [{\citenamefont {Chen}\ \emph {et~al.}(2023)\citenamefont {Chen}, \citenamefont {Huang}, \citenamefont {Jiang},\ and\ \citenamefont {Hu}}]{cyg}%
  \BibitemOpen
  \bibfield  {author} {\bibinfo {author} {\bibfnamefont {Y.}~\bibnamefont {Chen}}, \bibinfo {author} {\bibfnamefont {J.}~\bibnamefont {Huang}}, \bibinfo {author} {\bibfnamefont {K.}~\bibnamefont {Jiang}},\ and\ \bibinfo {author} {\bibfnamefont {J.}~\bibnamefont {Hu}},\ }\href {https://doi.org/https://doi.org/10.1016/j.scib.2023.11.032} {\bibfield  {journal} {\bibinfo  {journal} {Science Bulletin}\ }\textbf {\bibinfo {volume} {68}},\ \bibinfo {pages} {3165} (\bibinfo {year} {2023})}\BibitemShut {NoStop}%
\bibitem [{\citenamefont {Bergman}\ \emph {et~al.}(2008)\citenamefont {Bergman}, \citenamefont {Wu},\ and\ \citenamefont {Balents}}]{wu08}%
  \BibitemOpen
  \bibfield  {author} {\bibinfo {author} {\bibfnamefont {D.~L.}\ \bibnamefont {Bergman}}, \bibinfo {author} {\bibfnamefont {C.}~\bibnamefont {Wu}},\ and\ \bibinfo {author} {\bibfnamefont {L.}~\bibnamefont {Balents}},\ }\href {https://doi.org/10.1103/PhysRevB.78.125104} {\bibfield  {journal} {\bibinfo  {journal} {Phys. Rev. B}\ }\textbf {\bibinfo {volume} {78}},\ \bibinfo {pages} {125104} (\bibinfo {year} {2008})}\BibitemShut {NoStop}%
\bibitem [{\citenamefont {Rhim}\ and\ \citenamefont {Yang}(2019)}]{SFBprb}%
  \BibitemOpen
  \bibfield  {author} {\bibinfo {author} {\bibfnamefont {J.-W.}\ \bibnamefont {Rhim}}\ and\ \bibinfo {author} {\bibfnamefont {B.-J.}\ \bibnamefont {Yang}},\ }\href {https://doi.org/10.1103/PhysRevB.99.045107} {\bibfield  {journal} {\bibinfo  {journal} {Phys. Rev. B}\ }\textbf {\bibinfo {volume} {99}},\ \bibinfo {pages} {045107} (\bibinfo {year} {2019})}\BibitemShut {NoStop}%
\bibitem [{\citenamefont {Rhim}\ and\ \citenamefont {Yang}(2021)}]{SFBAPX}%
  \BibitemOpen
  \bibfield  {author} {\bibinfo {author} {\bibfnamefont {J.-W.}\ \bibnamefont {Rhim}}\ and\ \bibinfo {author} {\bibfnamefont {B.-J.}\ \bibnamefont {Yang}},\ }\href {https://doi.org/10.1080/23746149.2021.1901606} {\bibfield  {journal} {\bibinfo  {journal} {Advances in Physics: X}\ }\textbf {\bibinfo {volume} {6}},\ \bibinfo {pages} {1901606} (\bibinfo {year} {2021})}\BibitemShut {NoStop}%
\bibitem [{\citenamefont {Piéchon}\ \emph {et~al.}(2015)\citenamefont {Piéchon}, \citenamefont {Fuchs}, \citenamefont {Raoux},\ and\ \citenamefont {Montambaux}}]{frederic1}%
  \BibitemOpen
  \bibfield  {author} {\bibinfo {author} {\bibfnamefont {F.}~\bibnamefont {Piéchon}}, \bibinfo {author} {\bibfnamefont {J.-N.}\ \bibnamefont {Fuchs}}, \bibinfo {author} {\bibfnamefont {A.}~\bibnamefont {Raoux}},\ and\ \bibinfo {author} {\bibfnamefont {G.}~\bibnamefont {Montambaux}},\ }\href {https://doi.org/10.1088/1742-6596/603/1/012001} {\bibfield  {journal} {\bibinfo  {journal} {Journal of Physics: Conference Series}\ }\textbf {\bibinfo {volume} {603}},\ \bibinfo {pages} {012001} (\bibinfo {year} {2015})}\BibitemShut {NoStop}%
\bibitem [{\citenamefont {Pi\'echon}\ \emph {et~al.}(2016)\citenamefont {Pi\'echon}, \citenamefont {Raoux}, \citenamefont {Fuchs},\ and\ \citenamefont {Montambaux}}]{frederic2}%
  \BibitemOpen
  \bibfield  {author} {\bibinfo {author} {\bibfnamefont {F.}~\bibnamefont {Pi\'echon}}, \bibinfo {author} {\bibfnamefont {A.}~\bibnamefont {Raoux}}, \bibinfo {author} {\bibfnamefont {J.-N.}\ \bibnamefont {Fuchs}},\ and\ \bibinfo {author} {\bibfnamefont {G.}~\bibnamefont {Montambaux}},\ }\href {https://doi.org/10.1103/PhysRevB.94.134423} {\bibfield  {journal} {\bibinfo  {journal} {Phys. Rev. B}\ }\textbf {\bibinfo {volume} {94}},\ \bibinfo {pages} {134423} (\bibinfo {year} {2016})}\BibitemShut {NoStop}%
\bibitem [{\citenamefont {Graf}\ and\ \citenamefont {Pi\'echon}(2021)}]{frederic3}%
  \BibitemOpen
  \bibfield  {author} {\bibinfo {author} {\bibfnamefont {A.}~\bibnamefont {Graf}}\ and\ \bibinfo {author} {\bibfnamefont {F.}~\bibnamefont {Pi\'echon}},\ }\href {https://doi.org/10.1103/PhysRevB.104.195128} {\bibfield  {journal} {\bibinfo  {journal} {Phys. Rev. B}\ }\textbf {\bibinfo {volume} {104}},\ \bibinfo {pages} {195128} (\bibinfo {year} {2021})}\BibitemShut {NoStop}%
\bibitem [{\citenamefont {Raoux}\ \emph {et~al.}(2014)\citenamefont {Raoux}, \citenamefont {Morigi}, \citenamefont {Fuchs}, \citenamefont {Pi\'echon},\ and\ \citenamefont {Montambaux}}]{frederic4}%
  \BibitemOpen
  \bibfield  {author} {\bibinfo {author} {\bibfnamefont {A.}~\bibnamefont {Raoux}}, \bibinfo {author} {\bibfnamefont {M.}~\bibnamefont {Morigi}}, \bibinfo {author} {\bibfnamefont {J.-N.}\ \bibnamefont {Fuchs}}, \bibinfo {author} {\bibfnamefont {F.}~\bibnamefont {Pi\'echon}},\ and\ \bibinfo {author} {\bibfnamefont {G.}~\bibnamefont {Montambaux}},\ }\href {https://doi.org/10.1103/PhysRevLett.112.026402} {\bibfield  {journal} {\bibinfo  {journal} {Phys. Rev. Lett.}\ }\textbf {\bibinfo {volume} {112}},\ \bibinfo {pages} {026402} (\bibinfo {year} {2014})}\BibitemShut {NoStop}%
\bibitem [{\citenamefont {Vidal}\ \emph {et~al.}(2001)\citenamefont {Vidal}, \citenamefont {Butaud}, \citenamefont {Dou\ifmmode~\mbox{\c{c}}\else \c{c}\fi{}ot},\ and\ \citenamefont {Mosseri}}]{Julien1}%
  \BibitemOpen
  \bibfield  {author} {\bibinfo {author} {\bibfnamefont {J.}~\bibnamefont {Vidal}}, \bibinfo {author} {\bibfnamefont {P.}~\bibnamefont {Butaud}}, \bibinfo {author} {\bibfnamefont {B.}~\bibnamefont {Dou\ifmmode~\mbox{\c{c}}\else \c{c}\fi{}ot}},\ and\ \bibinfo {author} {\bibfnamefont {R.}~\bibnamefont {Mosseri}},\ }\href {https://doi.org/10.1103/PhysRevB.64.155306} {\bibfield  {journal} {\bibinfo  {journal} {Phys. Rev. B}\ }\textbf {\bibinfo {volume} {64}},\ \bibinfo {pages} {155306} (\bibinfo {year} {2001})}\BibitemShut {NoStop}%
\bibitem [{\citenamefont {Vidal}\ \emph {et~al.}(1998)\citenamefont {Vidal}, \citenamefont {Mosseri},\ and\ \citenamefont {Dou\ifmmode~\mbox{\c{c}}\else \c{c}\fi{}ot}}]{Julien2}%
  \BibitemOpen
  \bibfield  {author} {\bibinfo {author} {\bibfnamefont {J.}~\bibnamefont {Vidal}}, \bibinfo {author} {\bibfnamefont {R.}~\bibnamefont {Mosseri}},\ and\ \bibinfo {author} {\bibfnamefont {B.}~\bibnamefont {Dou\ifmmode~\mbox{\c{c}}\else \c{c}\fi{}ot}},\ }\href {https://doi.org/10.1103/PhysRevLett.81.5888} {\bibfield  {journal} {\bibinfo  {journal} {Phys. Rev. Lett.}\ }\textbf {\bibinfo {volume} {81}},\ \bibinfo {pages} {5888} (\bibinfo {year} {1998})}\BibitemShut {NoStop}%
\bibitem [{\citenamefont {Vidal}\ \emph {et~al.}(2000)\citenamefont {Vidal}, \citenamefont {Dou\ifmmode~\mbox{\c{c}}\else \c{c}\fi{}ot}, \citenamefont {Mosseri},\ and\ \citenamefont {Butaud}}]{Julien3}%
  \BibitemOpen
  \bibfield  {author} {\bibinfo {author} {\bibfnamefont {J.}~\bibnamefont {Vidal}}, \bibinfo {author} {\bibfnamefont {B.}~\bibnamefont {Dou\ifmmode~\mbox{\c{c}}\else \c{c}\fi{}ot}}, \bibinfo {author} {\bibfnamefont {R.}~\bibnamefont {Mosseri}},\ and\ \bibinfo {author} {\bibfnamefont {P.}~\bibnamefont {Butaud}},\ }\href {https://doi.org/10.1103/PhysRevLett.85.3906} {\bibfield  {journal} {\bibinfo  {journal} {Phys. Rev. Lett.}\ }\textbf {\bibinfo {volume} {85}},\ \bibinfo {pages} {3906} (\bibinfo {year} {2000})}\BibitemShut {NoStop}%
\bibitem [{\citenamefont {Dou\ifmmode~\mbox{\c{c}}\else \c{c}\fi{}ot}\ and\ \citenamefont {Vidal}(2002)}]{Julien4}%
  \BibitemOpen
  \bibfield  {author} {\bibinfo {author} {\bibfnamefont {B.}~\bibnamefont {Dou\ifmmode~\mbox{\c{c}}\else \c{c}\fi{}ot}}\ and\ \bibinfo {author} {\bibfnamefont {J.}~\bibnamefont {Vidal}},\ }\href {https://doi.org/10.1103/PhysRevLett.88.227005} {\bibfield  {journal} {\bibinfo  {journal} {Phys. Rev. Lett.}\ }\textbf {\bibinfo {volume} {88}},\ \bibinfo {pages} {227005} (\bibinfo {year} {2002})}\BibitemShut {NoStop}%
\bibitem [{\citenamefont {Dias}\ and\ \citenamefont {Gouveia}(2015)}]{Dias1}%
  \BibitemOpen
  \bibfield  {author} {\bibinfo {author} {\bibfnamefont {R.~G.}\ \bibnamefont {Dias}}\ and\ \bibinfo {author} {\bibfnamefont {J.~D.}\ \bibnamefont {Gouveia}},\ }\href {https://doi.org/10.1038/srep16852} {\bibfield  {journal} {\bibinfo  {journal} {Scientific Reports}\ }\textbf {\bibinfo {volume} {5}},\ \bibinfo {pages} {16852} (\bibinfo {year} {2015})}\BibitemShut {NoStop}%
\bibitem [{\citenamefont {Santos}\ and\ \citenamefont {Dias}(2020)}]{Dias2}%
  \BibitemOpen
  \bibfield  {author} {\bibinfo {author} {\bibfnamefont {F.~D.~R.}\ \bibnamefont {Santos}}\ and\ \bibinfo {author} {\bibfnamefont {R.~G.}\ \bibnamefont {Dias}},\ }\href {https://doi.org/10.1038/s41598-020-60975-7} {\bibfield  {journal} {\bibinfo  {journal} {Scientific Reports}\ }\textbf {\bibinfo {volume} {10}},\ \bibinfo {pages} {4532} (\bibinfo {year} {2020})}\BibitemShut {NoStop}%
\bibitem [{\citenamefont {Biswas}\ and\ \citenamefont {Chakrabarti}(2023)}]{referee1-1}%
  \BibitemOpen
  \bibfield  {author} {\bibinfo {author} {\bibfnamefont {S.}~\bibnamefont {Biswas}}\ and\ \bibinfo {author} {\bibfnamefont {A.}~\bibnamefont {Chakrabarti}},\ }\href {https://doi.org/https://doi.org/10.1016/j.physe.2023.115762} {\bibfield  {journal} {\bibinfo  {journal} {Physica E: Low-dimensional Systems and Nanostructures}\ }\textbf {\bibinfo {volume} {153}},\ \bibinfo {pages} {115762} (\bibinfo {year} {2023})}\BibitemShut {NoStop}%
\bibitem [{\citenamefont {Nandy}(2021)}]{referee1-2}%
  \BibitemOpen
  \bibfield  {author} {\bibinfo {author} {\bibfnamefont {A.}~\bibnamefont {Nandy}},\ }\href {https://doi.org/10.1088/1402-4896/abdcf6} {\bibfield  {journal} {\bibinfo  {journal} {Physica Scripta}\ }\textbf {\bibinfo {volume} {96}},\ \bibinfo {pages} {045802} (\bibinfo {year} {2021})}\BibitemShut {NoStop}%
\bibitem [{\citenamefont {Nandy}\ and\ \citenamefont {Chakrabarti}(2015)}]{referee1-3}%
  \BibitemOpen
  \bibfield  {author} {\bibinfo {author} {\bibfnamefont {A.}~\bibnamefont {Nandy}}\ and\ \bibinfo {author} {\bibfnamefont {A.}~\bibnamefont {Chakrabarti}},\ }\href {https://doi.org/https://doi.org/10.1016/j.physleta.2015.09.023} {\bibfield  {journal} {\bibinfo  {journal} {Physics Letters A}\ }\textbf {\bibinfo {volume} {379}},\ \bibinfo {pages} {2876} (\bibinfo {year} {2015})}\BibitemShut {NoStop}%
\bibitem [{\citenamefont {Nandy}\ \emph {et~al.}(2015)\citenamefont {Nandy}, \citenamefont {Pal},\ and\ \citenamefont {Chakrabarti}}]{referee1-4}%
  \BibitemOpen
  \bibfield  {author} {\bibinfo {author} {\bibfnamefont {A.}~\bibnamefont {Nandy}}, \bibinfo {author} {\bibfnamefont {B.}~\bibnamefont {Pal}},\ and\ \bibinfo {author} {\bibfnamefont {A.}~\bibnamefont {Chakrabarti}},\ }\href {https://doi.org/10.1088/0953-8984/27/12/125501} {\bibfield  {journal} {\bibinfo  {journal} {Journal of Physics: Condensed Matter}\ }\textbf {\bibinfo {volume} {27}},\ \bibinfo {pages} {125501} (\bibinfo {year} {2015})}\BibitemShut {NoStop}%
\bibitem [{\citenamefont {Pal}\ and\ \citenamefont {Saha}(2018)}]{referee1-5}%
  \BibitemOpen
  \bibfield  {author} {\bibinfo {author} {\bibfnamefont {B.}~\bibnamefont {Pal}}\ and\ \bibinfo {author} {\bibfnamefont {K.}~\bibnamefont {Saha}},\ }\href {https://doi.org/10.1103/PhysRevB.97.195101} {\bibfield  {journal} {\bibinfo  {journal} {Phys. Rev. B}\ }\textbf {\bibinfo {volume} {97}},\ \bibinfo {pages} {195101} (\bibinfo {year} {2018})}\BibitemShut {NoStop}%
\bibitem [{\citenamefont {Maimaiti}\ \emph {et~al.}(2019)\citenamefont {Maimaiti}, \citenamefont {Flach},\ and\ \citenamefont {Andreanov}}]{referee1-6-2}%
  \BibitemOpen
  \bibfield  {author} {\bibinfo {author} {\bibfnamefont {W.}~\bibnamefont {Maimaiti}}, \bibinfo {author} {\bibfnamefont {S.}~\bibnamefont {Flach}},\ and\ \bibinfo {author} {\bibfnamefont {A.}~\bibnamefont {Andreanov}},\ }\href {https://doi.org/10.1103/PhysRevB.99.125129} {\bibfield  {journal} {\bibinfo  {journal} {Phys. Rev. B}\ }\textbf {\bibinfo {volume} {99}},\ \bibinfo {pages} {125129} (\bibinfo {year} {2019})}\BibitemShut {NoStop}%
\bibitem [{\citenamefont {Maimaiti}\ \emph {et~al.}(2021)\citenamefont {Maimaiti}, \citenamefont {Andreanov},\ and\ \citenamefont {Flach}}]{referee1-7-2}%
  \BibitemOpen
  \bibfield  {author} {\bibinfo {author} {\bibfnamefont {W.}~\bibnamefont {Maimaiti}}, \bibinfo {author} {\bibfnamefont {A.}~\bibnamefont {Andreanov}},\ and\ \bibinfo {author} {\bibfnamefont {S.}~\bibnamefont {Flach}},\ }\href {https://doi.org/10.1103/PhysRevB.103.165116} {\bibfield  {journal} {\bibinfo  {journal} {Phys. Rev. B}\ }\textbf {\bibinfo {volume} {103}},\ \bibinfo {pages} {165116} (\bibinfo {year} {2021})}\BibitemShut {NoStop}%
\bibitem [{\citenamefont {Flach}\ \emph {et~al.}(2014)\citenamefont {Flach}, \citenamefont {Leykam}, \citenamefont {Bodyfelt}, \citenamefont {Matthies},\ and\ \citenamefont {Desyatnikov}}]{CLS0}%
  \BibitemOpen
  \bibfield  {author} {\bibinfo {author} {\bibfnamefont {S.}~\bibnamefont {Flach}}, \bibinfo {author} {\bibfnamefont {D.}~\bibnamefont {Leykam}}, \bibinfo {author} {\bibfnamefont {J.~D.}\ \bibnamefont {Bodyfelt}}, \bibinfo {author} {\bibfnamefont {P.}~\bibnamefont {Matthies}},\ and\ \bibinfo {author} {\bibfnamefont {A.~S.}\ \bibnamefont {Desyatnikov}},\ }\href {https://doi.org/10.1209/0295-5075/105/30001} {\bibfield  {journal} {\bibinfo  {journal} {Europhysics Letters}\ }\textbf {\bibinfo {volume} {105}},\ \bibinfo {pages} {30001} (\bibinfo {year} {2014})}\BibitemShut {NoStop}%
\bibitem [{\citenamefont {R\"ontgen}\ \emph {et~al.}(2018)\citenamefont {R\"ontgen}, \citenamefont {Morfonios},\ and\ \citenamefont {Schmelcher}}]{CLS2}%
  \BibitemOpen
  \bibfield  {author} {\bibinfo {author} {\bibfnamefont {M.}~\bibnamefont {R\"ontgen}}, \bibinfo {author} {\bibfnamefont {C.~V.}\ \bibnamefont {Morfonios}},\ and\ \bibinfo {author} {\bibfnamefont {P.}~\bibnamefont {Schmelcher}},\ }\href {https://doi.org/10.1103/PhysRevB.97.035161} {\bibfield  {journal} {\bibinfo  {journal} {Phys. Rev. B}\ }\textbf {\bibinfo {volume} {97}},\ \bibinfo {pages} {035161} (\bibinfo {year} {2018})}\BibitemShut {NoStop}%
\bibitem [{\citenamefont {He}\ \emph {et~al.}(2025{\natexlab{a}})\citenamefont {He}, \citenamefont {Li}, \citenamefont {Zhang}, \citenamefont {Fu}, \citenamefont {Li},\ and\ \citenamefont {Zhong}}]{he1}%
  \BibitemOpen
  \bibfield  {author} {\bibinfo {author} {\bibfnamefont {C.}~\bibnamefont {He}}, \bibinfo {author} {\bibfnamefont {S.}~\bibnamefont {Li}}, \bibinfo {author} {\bibfnamefont {Y.}~\bibnamefont {Zhang}}, \bibinfo {author} {\bibfnamefont {Z.}~\bibnamefont {Fu}}, \bibinfo {author} {\bibfnamefont {J.}~\bibnamefont {Li}},\ and\ \bibinfo {author} {\bibfnamefont {J.}~\bibnamefont {Zhong}},\ }\href {https://doi.org/10.1103/PhysRevB.111.L081404} {\bibfield  {journal} {\bibinfo  {journal} {Phys. Rev. B}\ }\textbf {\bibinfo {volume} {111}},\ \bibinfo {pages} {L081404} (\bibinfo {year} {2025}{\natexlab{a}})}\BibitemShut {NoStop}%
\bibitem [{\citenamefont {He}\ \emph {et~al.}(2025{\natexlab{b}})\citenamefont {He}, \citenamefont {Liao}, \citenamefont {Ouyang}, \citenamefont {Zhang}, \citenamefont {Xiang},\ and\ \citenamefont {Zhong}}]{he2}%
  \BibitemOpen
  \bibfield  {author} {\bibinfo {author} {\bibfnamefont {C.}~\bibnamefont {He}}, \bibinfo {author} {\bibfnamefont {Y.}~\bibnamefont {Liao}}, \bibinfo {author} {\bibfnamefont {T.}~\bibnamefont {Ouyang}}, \bibinfo {author} {\bibfnamefont {H.}~\bibnamefont {Zhang}}, \bibinfo {author} {\bibfnamefont {H.}~\bibnamefont {Xiang}},\ and\ \bibinfo {author} {\bibfnamefont {J.}~\bibnamefont {Zhong}},\ }\href {https://doi.org/https://doi.org/10.1016/j.fmre.2023.12.001} {\bibfield  {journal} {\bibinfo  {journal} {Fundamental Research}\ }\textbf {\bibinfo {volume} {5}},\ \bibinfo {pages} {138} (\bibinfo {year} {2025}{\natexlab{b}})}\BibitemShut {NoStop}%
\bibitem [{\citenamefont {Liu}\ \emph {et~al.}(2014)\citenamefont {Liu}, \citenamefont {Liu},\ and\ \citenamefont {Wu}}]{LiuWu}%
  \BibitemOpen
  \bibfield  {author} {\bibinfo {author} {\bibfnamefont {Z.}~\bibnamefont {Liu}}, \bibinfo {author} {\bibfnamefont {F.}~\bibnamefont {Liu}},\ and\ \bibinfo {author} {\bibfnamefont {Y.-S.}\ \bibnamefont {Wu}},\ }\href {https://doi.org/10.1088/1674-1056/23/7/077308} {\bibfield  {journal} {\bibinfo  {journal} {Chinese Physics B}\ }\textbf {\bibinfo {volume} {23}},\ \bibinfo {pages} {077308} (\bibinfo {year} {2014})}\BibitemShut {NoStop}%
\bibitem [{\citenamefont {de~Jesús Espinosa-Champo}\ and\ \citenamefont {Naumis}(2024)}]{holed1}%
  \BibitemOpen
  \bibfield  {author} {\bibinfo {author} {\bibfnamefont {A.}~\bibnamefont {de~Jesús Espinosa-Champo}}\ and\ \bibinfo {author} {\bibfnamefont {G.~G.}\ \bibnamefont {Naumis}},\ }\href {https://doi.org/10.1088/1361-648X/ad39be} {\bibfield  {journal} {\bibinfo  {journal} {Journal of Physics: Condensed Matter}\ }\textbf {\bibinfo {volume} {36}},\ \bibinfo {pages} {275703} (\bibinfo {year} {2024})}\BibitemShut {NoStop}%
\bibitem [{\citenamefont {Sethi}\ \emph {et~al.}(2021)\citenamefont {Sethi}, \citenamefont {Zhou}, \citenamefont {Zhu}, \citenamefont {Yang},\ and\ \citenamefont {Liu}}]{YYFBliu}%
  \BibitemOpen
  \bibfield  {author} {\bibinfo {author} {\bibfnamefont {G.}~\bibnamefont {Sethi}}, \bibinfo {author} {\bibfnamefont {Y.}~\bibnamefont {Zhou}}, \bibinfo {author} {\bibfnamefont {L.}~\bibnamefont {Zhu}}, \bibinfo {author} {\bibfnamefont {L.}~\bibnamefont {Yang}},\ and\ \bibinfo {author} {\bibfnamefont {F.}~\bibnamefont {Liu}},\ }\href {https://doi.org/10.1103/PhysRevLett.126.196403} {\bibfield  {journal} {\bibinfo  {journal} {Phys. Rev. Lett.}\ }\textbf {\bibinfo {volume} {126}},\ \bibinfo {pages} {196403} (\bibinfo {year} {2021})}\BibitemShut {NoStop}%
\bibitem [{\citenamefont {Kim}\ and\ \citenamefont {Liu}(2023)}]{liufeng1}%
  \BibitemOpen
  \bibfield  {author} {\bibinfo {author} {\bibfnamefont {D.}~\bibnamefont {Kim}}\ and\ \bibinfo {author} {\bibfnamefont {F.}~\bibnamefont {Liu}},\ }\href {https://doi.org/10.1103/PhysRevB.107.205130} {\bibfield  {journal} {\bibinfo  {journal} {Phys. Rev. B}\ }\textbf {\bibinfo {volume} {107}},\ \bibinfo {pages} {205130} (\bibinfo {year} {2023})}\BibitemShut {NoStop}%
\bibitem [{\citenamefont {Pan}\ \emph {et~al.}(2023)\citenamefont {Pan}, \citenamefont {Zhang}, \citenamefont {Zhou}, \citenamefont {Wang}, \citenamefont {Bian}, \citenamefont {Liu}, \citenamefont {Wang}, \citenamefont {Li}, \citenamefont {Chen}, \citenamefont {Lei}, \citenamefont {Li}, \citenamefont {Cheng}, \citenamefont {Shao}, \citenamefont {Ding}, \citenamefont {Gao}, \citenamefont {Li},\ and\ \citenamefont {Liu}}]{liufeng2}%
  \BibitemOpen
  \bibfield  {author} {\bibinfo {author} {\bibfnamefont {M.}~\bibnamefont {Pan}}, \bibinfo {author} {\bibfnamefont {X.}~\bibnamefont {Zhang}}, \bibinfo {author} {\bibfnamefont {Y.}~\bibnamefont {Zhou}}, \bibinfo {author} {\bibfnamefont {P.}~\bibnamefont {Wang}}, \bibinfo {author} {\bibfnamefont {Q.}~\bibnamefont {Bian}}, \bibinfo {author} {\bibfnamefont {H.}~\bibnamefont {Liu}}, \bibinfo {author} {\bibfnamefont {X.}~\bibnamefont {Wang}}, \bibinfo {author} {\bibfnamefont {X.}~\bibnamefont {Li}}, \bibinfo {author} {\bibfnamefont {A.}~\bibnamefont {Chen}}, \bibinfo {author} {\bibfnamefont {X.}~\bibnamefont {Lei}}, \bibinfo {author} {\bibfnamefont {S.}~\bibnamefont {Li}}, \bibinfo {author} {\bibfnamefont {Z.}~\bibnamefont {Cheng}}, \bibinfo {author} {\bibfnamefont {Z.}~\bibnamefont {Shao}}, \bibinfo {author} {\bibfnamefont {H.}~\bibnamefont {Ding}}, \bibinfo {author} {\bibfnamefont {J.}~\bibnamefont {Gao}}, \bibinfo {author} {\bibfnamefont {F.}~\bibnamefont {Li}},\ and\ \bibinfo {author} {\bibfnamefont
  {F.}~\bibnamefont {Liu}},\ }\href {https://doi.org/10.1103/PhysRevLett.130.036203} {\bibfield  {journal} {\bibinfo  {journal} {Phys. Rev. Lett.}\ }\textbf {\bibinfo {volume} {130}},\ \bibinfo {pages} {036203} (\bibinfo {year} {2023})}\BibitemShut {NoStop}%
\bibitem [{\citenamefont {Lieb}(1989)}]{Lieb}%
  \BibitemOpen
  \bibfield  {author} {\bibinfo {author} {\bibfnamefont {E.~H.}\ \bibnamefont {Lieb}},\ }\href {https://doi.org/10.1103/PhysRevLett.62.1201} {\bibfield  {journal} {\bibinfo  {journal} {Phys. Rev. Lett.}\ }\textbf {\bibinfo {volume} {62}},\ \bibinfo {pages} {1201} (\bibinfo {year} {1989})}\BibitemShut {NoStop}%
\bibitem [{\citenamefont {Jiang}\ \emph {et~al.}(2019)\citenamefont {Jiang}, \citenamefont {Kang}, \citenamefont {Huang}, \citenamefont {Xu}, \citenamefont {Low},\ and\ \citenamefont {Liu}}]{LiuLieb}%
  \BibitemOpen
  \bibfield  {author} {\bibinfo {author} {\bibfnamefont {W.}~\bibnamefont {Jiang}}, \bibinfo {author} {\bibfnamefont {M.}~\bibnamefont {Kang}}, \bibinfo {author} {\bibfnamefont {H.}~\bibnamefont {Huang}}, \bibinfo {author} {\bibfnamefont {H.}~\bibnamefont {Xu}}, \bibinfo {author} {\bibfnamefont {T.}~\bibnamefont {Low}},\ and\ \bibinfo {author} {\bibfnamefont {F.}~\bibnamefont {Liu}},\ }\href {https://doi.org/10.1103/PhysRevB.99.125131} {\bibfield  {journal} {\bibinfo  {journal} {Phys. Rev. B}\ }\textbf {\bibinfo {volume} {99}},\ \bibinfo {pages} {125131} (\bibinfo {year} {2019})}\BibitemShut {NoStop}%
\bibitem [{\citenamefont {Xia}\ \emph {et~al.}(2018)\citenamefont {Xia}, \citenamefont {Ramachandran}, \citenamefont {Xia}, \citenamefont {Li}, \citenamefont {Liu}, \citenamefont {Tang}, \citenamefont {Hu}, \citenamefont {Song}, \citenamefont {Xu}, \citenamefont {Leykam}, \citenamefont {Flach},\ and\ \citenamefont {Chen}}]{linestate0}%
  \BibitemOpen
  \bibfield  {author} {\bibinfo {author} {\bibfnamefont {S.}~\bibnamefont {Xia}}, \bibinfo {author} {\bibfnamefont {A.}~\bibnamefont {Ramachandran}}, \bibinfo {author} {\bibfnamefont {S.}~\bibnamefont {Xia}}, \bibinfo {author} {\bibfnamefont {D.}~\bibnamefont {Li}}, \bibinfo {author} {\bibfnamefont {X.}~\bibnamefont {Liu}}, \bibinfo {author} {\bibfnamefont {L.}~\bibnamefont {Tang}}, \bibinfo {author} {\bibfnamefont {Y.}~\bibnamefont {Hu}}, \bibinfo {author} {\bibfnamefont {D.}~\bibnamefont {Song}}, \bibinfo {author} {\bibfnamefont {J.}~\bibnamefont {Xu}}, \bibinfo {author} {\bibfnamefont {D.}~\bibnamefont {Leykam}}, \bibinfo {author} {\bibfnamefont {S.}~\bibnamefont {Flach}},\ and\ \bibinfo {author} {\bibfnamefont {Z.}~\bibnamefont {Chen}},\ }\href {https://doi.org/10.1103/PhysRevLett.121.263902} {\bibfield  {journal} {\bibinfo  {journal} {Phys. Rev. Lett.}\ }\textbf {\bibinfo {volume} {121}},\ \bibinfo {pages} {263902} (\bibinfo {year} {2018})}\BibitemShut {NoStop}%
\bibitem [{\citenamefont {Yan}\ \emph {et~al.}(2020)\citenamefont {Yan}, \citenamefont {Zhong}, \citenamefont {Song}, \citenamefont {Zhang}, \citenamefont {Xia}, \citenamefont {Tang}, \citenamefont {Leykam},\ and\ \citenamefont {Chen}}]{linestate1}%
  \BibitemOpen
  \bibfield  {author} {\bibinfo {author} {\bibfnamefont {W.}~\bibnamefont {Yan}}, \bibinfo {author} {\bibfnamefont {H.}~\bibnamefont {Zhong}}, \bibinfo {author} {\bibfnamefont {D.}~\bibnamefont {Song}}, \bibinfo {author} {\bibfnamefont {Y.}~\bibnamefont {Zhang}}, \bibinfo {author} {\bibfnamefont {S.}~\bibnamefont {Xia}}, \bibinfo {author} {\bibfnamefont {L.}~\bibnamefont {Tang}}, \bibinfo {author} {\bibfnamefont {D.}~\bibnamefont {Leykam}},\ and\ \bibinfo {author} {\bibfnamefont {Z.}~\bibnamefont {Chen}},\ }\href {https://doi.org/https://doi.org/10.1002/adom.201902174} {\bibfield  {journal} {\bibinfo  {journal} {Advanced Optical Materials}\ }\textbf {\bibinfo {volume} {8}},\ \bibinfo {pages} {1902174} (\bibinfo {year} {2020})}\BibitemShut {NoStop}%
\bibitem [{\citenamefont {Amo}\ \emph {et~al.}(2021)\citenamefont {Amo}, \citenamefont {Ozawa},\ and\ \citenamefont {El-Ganainy}}]{linestate2}%
  \BibitemOpen
  \bibfield  {author} {\bibinfo {author} {\bibfnamefont {A.}~\bibnamefont {Amo}}, \bibinfo {author} {\bibfnamefont {T.}~\bibnamefont {Ozawa}},\ and\ \bibinfo {author} {\bibfnamefont {R.}~\bibnamefont {El-Ganainy}},\ }\href {https://doi.org/10.1364/OME.427194} {\bibfield  {journal} {\bibinfo  {journal} {Opt. Mater. Express}\ }\textbf {\bibinfo {volume} {11}},\ \bibinfo {pages} {1410} (\bibinfo {year} {2021})}\BibitemShut {NoStop}%
\bibitem [{\citenamefont {Song}\ \emph {et~al.}(2023)\citenamefont {Song}, \citenamefont {Xie}, \citenamefont {Xia}, \citenamefont {Tang}, \citenamefont {Song}, \citenamefont {Rhim},\ and\ \citenamefont {Chen}}]{opticalSFB1}%
  \BibitemOpen
  \bibfield  {author} {\bibinfo {author} {\bibfnamefont {L.}~\bibnamefont {Song}}, \bibinfo {author} {\bibfnamefont {Y.}~\bibnamefont {Xie}}, \bibinfo {author} {\bibfnamefont {S.}~\bibnamefont {Xia}}, \bibinfo {author} {\bibfnamefont {L.}~\bibnamefont {Tang}}, \bibinfo {author} {\bibfnamefont {D.}~\bibnamefont {Song}}, \bibinfo {author} {\bibfnamefont {J.-W.}\ \bibnamefont {Rhim}},\ and\ \bibinfo {author} {\bibfnamefont {Z.}~\bibnamefont {Chen}},\ }\href {https://doi.org/https://doi.org/10.1002/lpor.202200315} {\bibfield  {journal} {\bibinfo  {journal} {Laser \& Photonics Reviews}\ }\textbf {\bibinfo {volume} {17}},\ \bibinfo {pages} {2200315} (\bibinfo {year} {2023})}\BibitemShut {NoStop}%
\bibitem [{\citenamefont {Menz}\ \emph {et~al.}(2023)\citenamefont {Menz}, \citenamefont {Hanafi}, \citenamefont {Imbrock},\ and\ \citenamefont {Denz}}]{opticalSFB2}%
  \BibitemOpen
  \bibfield  {author} {\bibinfo {author} {\bibfnamefont {P.}~\bibnamefont {Menz}}, \bibinfo {author} {\bibfnamefont {H.}~\bibnamefont {Hanafi}}, \bibinfo {author} {\bibfnamefont {J.}~\bibnamefont {Imbrock}},\ and\ \bibinfo {author} {\bibfnamefont {C.}~\bibnamefont {Denz}},\ }\href {https://doi.org/doi:10.1515/nanoph-2023-0222} {\bibfield  {journal} {\bibinfo  {journal} {Nanophotonics}\ }\textbf {\bibinfo {volume} {12}},\ \bibinfo {pages} {3409} (\bibinfo {year} {2023})}\BibitemShut {NoStop}%
\bibitem [{\citenamefont {Song}\ \emph {et~al.}(2025)\citenamefont {Song}, \citenamefont {Gao}, \citenamefont {Xia}, \citenamefont {Liang}, \citenamefont {Tang}, \citenamefont {Song}, \citenamefont {Leykam},\ and\ \citenamefont {Chen}}]{opticalSFB3}%
  \BibitemOpen
  \bibfield  {author} {\bibinfo {author} {\bibfnamefont {L.}~\bibnamefont {Song}}, \bibinfo {author} {\bibfnamefont {S.}~\bibnamefont {Gao}}, \bibinfo {author} {\bibfnamefont {S.}~\bibnamefont {Xia}}, \bibinfo {author} {\bibfnamefont {Y.}~\bibnamefont {Liang}}, \bibinfo {author} {\bibfnamefont {L.}~\bibnamefont {Tang}}, \bibinfo {author} {\bibfnamefont {D.}~\bibnamefont {Song}}, \bibinfo {author} {\bibfnamefont {D.}~\bibnamefont {Leykam}},\ and\ \bibinfo {author} {\bibfnamefont {Z.}~\bibnamefont {Chen}},\ }\href {https://doi.org/10.1103/PhysRevLett.134.063803} {\bibfield  {journal} {\bibinfo  {journal} {Phys. Rev. Lett.}\ }\textbf {\bibinfo {volume} {134}},\ \bibinfo {pages} {063803} (\bibinfo {year} {2025})}\BibitemShut {NoStop}%
\bibitem [{\citenamefont {Tsai}\ \emph {et~al.}(2015)\citenamefont {Tsai}, \citenamefont {Fang}, \citenamefont {Yao},\ and\ \citenamefont {Hu}}]{Tsai_2015}%
  \BibitemOpen
  \bibfield  {author} {\bibinfo {author} {\bibfnamefont {W.-F.}\ \bibnamefont {Tsai}}, \bibinfo {author} {\bibfnamefont {C.}~\bibnamefont {Fang}}, \bibinfo {author} {\bibfnamefont {H.}~\bibnamefont {Yao}},\ and\ \bibinfo {author} {\bibfnamefont {J.}~\bibnamefont {Hu}},\ }\href {https://doi.org/10.1088/1367-2630/17/5/055016} {\bibfield  {journal} {\bibinfo  {journal} {New Journal of Physics}\ }\textbf {\bibinfo {volume} {17}},\ \bibinfo {pages} {055016} (\bibinfo {year} {2015})}\BibitemShut {NoStop}%
\bibitem [{\citenamefont {Jaworowski}\ \emph {et~al.}(2015)\citenamefont {Jaworowski}, \citenamefont {Manolescu},\ and\ \citenamefont {Potasz}}]{FCILieb}%
  \BibitemOpen
  \bibfield  {author} {\bibinfo {author} {\bibfnamefont {B.}~\bibnamefont {Jaworowski}}, \bibinfo {author} {\bibfnamefont {A.}~\bibnamefont {Manolescu}},\ and\ \bibinfo {author} {\bibfnamefont {P.}~\bibnamefont {Potasz}},\ }\href {https://doi.org/10.1103/PhysRevB.92.245119} {\bibfield  {journal} {\bibinfo  {journal} {Phys. Rev. B}\ }\textbf {\bibinfo {volume} {92}},\ \bibinfo {pages} {245119} (\bibinfo {year} {2015})}\BibitemShut {NoStop}%
\bibitem [{\citenamefont {Chen}\ \emph {et~al.}()\citenamefont {Chen}, \citenamefont {Yu}, \citenamefont {Huang}, \citenamefont {Zheng},\ and\ \citenamefont {Hu}}]{data-availbale}%
  \BibitemOpen
  \bibfield  {author} {\bibinfo {author} {\bibfnamefont {Y.}~\bibnamefont {Chen}}, \bibinfo {author} {\bibfnamefont {H.}~\bibnamefont {Yu}}, \bibinfo {author} {\bibfnamefont {Y.-P.}\ \bibnamefont {Huang}}, \bibinfo {author} {\bibfnamefont {Z.-Y.}\ \bibnamefont {Zheng}},\ and\ \bibinfo {author} {\bibfnamefont {J.}~\bibnamefont {Hu}},\ }\href@noop {} {\bibinfo  {journal} {Data available in zenodo for publication}\ }\BibitemShut {NoStop}%
\bibitem [{\citenamefont {Latorre}\ \emph {et~al.}(2004)\citenamefont {Latorre}, \citenamefont {Rico},\ and\ \citenamefont {Vidal}}]{vidal}%
  \BibitemOpen
\bibfield  {journal} {  }\bibfield  {author} {\bibinfo {author} {\bibfnamefont {J.~I.}\ \bibnamefont {Latorre}}, \bibinfo {author} {\bibfnamefont {E.}~\bibnamefont {Rico}},\ and\ \bibinfo {author} {\bibfnamefont {G.}~\bibnamefont {Vidal}},\ }\href {https://arxiv.org/abs/quant-ph/0304098} {\bibinfo {title} {Ground state entanglement in quantum spin chains}} (\bibinfo {year} {2004})\BibitemShut {NoStop}%
\bibitem [{\citenamefont {Jin}\ and\ \citenamefont {Korepin}(2004)}]{Jin2004}%
  \BibitemOpen
  \bibfield  {author} {\bibinfo {author} {\bibfnamefont {B.-Q.}\ \bibnamefont {Jin}}\ and\ \bibinfo {author} {\bibfnamefont {V.~E.}\ \bibnamefont {Korepin}},\ }\href {https://doi.org/10.1023/B:JOSS.0000037230.37166.42} {\bibfield  {journal} {\bibinfo  {journal} {Journal of Statistical Physics}\ }\textbf {\bibinfo {volume} {116}},\ \bibinfo {pages} {79} (\bibinfo {year} {2004})}\BibitemShut {NoStop}%
\bibitem [{\citenamefont {Vidal}\ \emph {et~al.}(2003)\citenamefont {Vidal}, \citenamefont {Latorre}, \citenamefont {Rico},\ and\ \citenamefont {Kitaev}}]{vidal2}%
  \BibitemOpen
  \bibfield  {author} {\bibinfo {author} {\bibfnamefont {G.}~\bibnamefont {Vidal}}, \bibinfo {author} {\bibfnamefont {J.~I.}\ \bibnamefont {Latorre}}, \bibinfo {author} {\bibfnamefont {E.}~\bibnamefont {Rico}},\ and\ \bibinfo {author} {\bibfnamefont {A.}~\bibnamefont {Kitaev}},\ }\href {https://doi.org/10.1103/PhysRevLett.90.227902} {\bibfield  {journal} {\bibinfo  {journal} {Phys. Rev. Lett.}\ }\textbf {\bibinfo {volume} {90}},\ \bibinfo {pages} {227902} (\bibinfo {year} {2003})}\BibitemShut {NoStop}%
\end{thebibliography}%

%\color bib list

%%%%%%%%%% Merge with supplemental materials %%%%%%%%%%
%\pagebreak
\newpage
\clearpage
\onecolumngrid
\begin{center}
\textbf{\large Supplemental Material for ''Topological Degeneracy Induced Flat Bands in two-Dimensional Holed Systems''}
\end{center}

\author{Yuge Chen}
\thanks{These three authors contributed equally}
\affiliation{Institute for Quantum Science and Technology, Shanghai University, Shanghai 200444, China}
\affiliation{Beijing National Laboratory for Condensed Matter Physics and Institute of Physics,
	Chinese Academy of Sciences, Beijing 100190, China}

\author{Hui Yu}
\thanks{These three authors contributed equally}
\affiliation{Beijing National Laboratory for Condensed Matter Physics and Institute of Physics,
	Chinese Academy of Sciences, Beijing 100190, China}
	
\author{Yun-Peng Huang}
\thanks{These three authors contributed equally}
\affiliation{Beijing National Laboratory for Condensed Matter Physics and Institute of Physics,
	Chinese Academy of Sciences, Beijing 100190, China}
 
\author{Zhen-Yu Zheng}
\affiliation{Beijing National Laboratory for Condensed Matter Physics and Institute of Physics,
	Chinese Academy of Sciences, Beijing 100190, China}

\author{Jiangping Hu}
\email{jphu@iphy.ac.cn}
\affiliation{Beijing National Laboratory for Condensed Matter Physics and Institute of Physics,
	Chinese Academy of Sciences, Beijing 100190, China}
\affiliation{Kavli Institute of Theoretical Sciences, University of Chinese Academy of Sciences,
	Beijing, 100190, China}
	 \affiliation{New Cornerstone Science Laboratory, 
	Beijing, 100190, China}
\maketitle

\section{Entanglement property of Lieb Lattice}
Although singular flat band systems exhibit topological degeneracy, they do not support fractional statistics, distinguishing them from systems with intrinsic topological order. This distinction raises an intriguing question: what are the entanglement properties of systems that possess topological degeneracy but do not qualify as topological ordered systems? To address this question, we revisit the Lieb lattice and calculate its entanglement entropy for different Fermi surfaces. Below, we outline the key steps of the computation for clarity, while the full derivation and technical details are provided in \cite{vidal}. 

\begin{figure}
\centering
  \includegraphics[width=11.5cm]{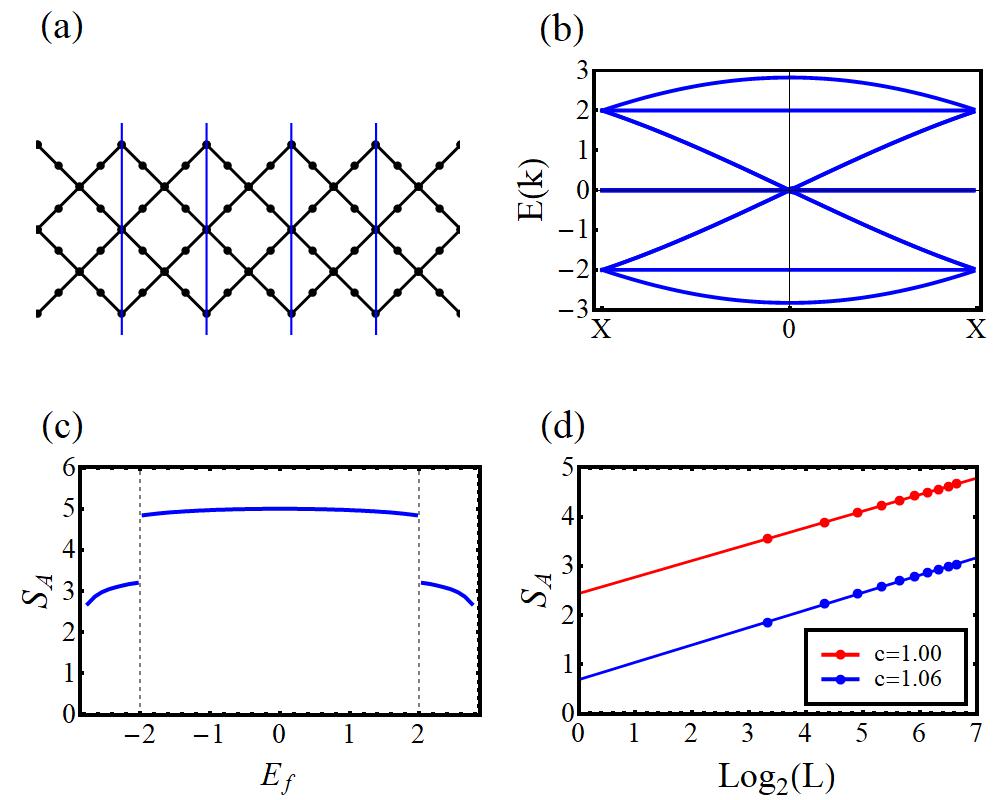}\\
  \caption{(a) A segment of the Lieb lattice model with periodic boundary conditions. The full system consists of 2 periods in the vertical direction and $L$ periods in the horizontal direction. The region enclosed by the two blue lines represent a unit cell. (b) The energy spectrum of the system (a). The degeneracies of the energy bands, ordered from top to bottom, are 1, 2, 1, 4, 1, 2, and 1, respectively. (c) The entanglement entropy of subsystem $A$ as a function of the Fermi level with $L=100$. (d) The entanglement entropy of subsystem $A$ as a function of the system's length $L$ at two different Fermi energies. Red dots: $E_f=-2.01t$. Blue dots: $E_f=-1.99t$. The colored dots represent numerical calculations, while the solid lines correspond to the fitting using Eq.~\ref{CFTentropy}, with $s=2.44$ and $s=0.69$ in both cases.}\label{figentangle}
\end{figure}

Here we consider the Lieb lattice with periodic boundary condition as a quasi-one-dimensional system, as depicted in Fig.~\ref{figentangle}(a). Given that the van Hove singularities of the Lieb lattice are located at the K point, we enlarge the unit cell by $\sqrt{2}\times\sqrt{2}$. The Hamiltonian of the system is identical to one descirbed in main text, with the chemical potential set to zero for simplicity. First, we diagonalize the Hamiltonian to obtain its complete set of eigenstates and corresponding eigenenergies. As reflected in Fig.~\ref{figentangle}(b), two flat bands emerge from van Hove singularities at $E_f=\pm 2t$. Then we construct the correlation matrix $\Gamma_{A}$ for subsystem A, which is defined as follows:

\begin{eqnarray}
\Gamma ^A=\left[
\begin{array}{cccc}
 \Gamma _{1,1} & \Gamma _{1,2} & \cdots  & \Gamma _{1,n} \\
 \Gamma _{2,1} & \Gamma _{2,2} & \cdots  & \Gamma _{2,n} \\
 \vdots  & \vdots  & \ddots & \vdots  \\
 \Gamma _{n,1} & \Gamma _{n,2} & \cdots  & \Gamma _{n,n} \\
\end{array}
\right]
\end{eqnarray}
Here, $\Gamma_{i,j}^{A}=\left<c_i^{\dagger }c_j\right>$ represents the expectation value of electron correlation function between the $i$-th and $j$-th lattice sites in region A to the ground state of the system. Additionally, $n$ is the total number of lattice sites within subsystem $A$. Next we diagonalize the correlation matrix $\Gamma^{A}$ to obtain its eigenvalues $\lambda_{i}$. Using these eigenvalues, the entanglement entropy $S_{A}$ of subsystem A can be calculated as follows:

\begin{equation}
    S_A=\sum _{i=1}^n \lambda _i Log_2\left(\lambda _i\right)+\left(1-\lambda _i\right)Log_2\left(1-\lambda _i\right)
\end{equation}

We set the length of subsystem A to be half of the entire system's length $L$. As indicated in Fig.~\ref{figentangle}(c), we calculate  $S_A$ by varying Fermi surfaces of the system and observe the gapes at van Hove singularity. We further investigate the behavior of the entanglement entropy $S_A$ as a function of the system size for two Fermi surfaces positioned on either side of the van Hove singularity. In both cases, $S_A$ exhibits a logarithmic scaling with system size, consistent with the behavior predicted by 1+1 dimensional conformal field theory  \cite{Jin2004}:

\begin{equation}\label{CFTentropy}
S_{A} =\frac{c}{3} Log_2(L)+s
\end{equation}
where $c$ is the central charge, which in our case equals $1$. This result implies that when the Fermi level lies either above or below the van Hove singularity of the Lieb lattice, the entanglement entropy of our model follows the same scaling form as that of a one-dimensional free massless complex fermion. Notably, this behavior is analogous to the entanglement structure observed in the XX and XXZ spin chains \cite{vidal,vidal2}. Furthermore, the positive value of $s$ in both scenarios indicate that the system does not exhibit topological order.
%\begin{center}
%\textbf{\large Supplemental Material: Flat bands from holed two dimension van Hove systems}
%\end{center}

%%%%%%%%%% Merge with supplemental materials %%%%%%%%%%

%%%%%%%%%% Prefix a "S" to all equations, figures, tables and reset the counter %%%%%%%%%%
%\setcounter{equation}{0}
%\setcounter{figure}{0}
%\setcounter{table}{0}
%\setcounter{page}{1}
%\makeatletter
%\renewcommand{\theequation}{S\arabic{equation}}
%\renewcommand{\thefigure}{S\arabic{figure}}
%\renewcommand{\thetable}{S\arabic{table}}
%\renewcommand{\bibnumfmt}[1]{[S#1]}
%\renewcommand{\citenumfont}[1]{S#1}
%%%%%%%%%% Prefix a "S" to all equations, figures, tables and reset the counter %%%%%%%%%%

%\twocolumngrid

%\subsection{short introduction about line states in holed honeycomb, triangle and kagome lattice}
%Like square lattice, other perfect nesting van Hove systems also exhibit loop excitations, as shown in the figure. For the wave functions excited by these loops, we can also design similar boundary conditions to protect the line excitation under open boundary conditions, as shown in the figure. These line excitations have topological properties similar to those in perforated square grids. As long as the boundaries of two holes are matched, no matter how far apart they are, there will always be an excitation in the system that connects the two boundaries. Moreover, line excitation is only sensitive to disturbances on the connecting paths in the system.

%\subsection{frustrated holed square lattice}

\end{document}